UNIVERSITY OF CALIFORNIA, MERCED

**Code Generation Techniques for Raw Data Processing**

A thesis submitted in partial satisfaction of the

requirements for the degree

Master of Science

in

Electrical Engineering and Computer Science

by

Xin Zhang

Committee in charge:

    Prof. Florin Rusu, Chair
    Prof. Dong Li
    Prof. Sungjin Im

2017

The thesis of Xin Zhang is approved, and it is acceptable in quality and form for publication on microfilm and electronically:

---

---

---

<div align="right">Chair</div>

University of California, Merced

2017

# TABLE OF CONTENTS

# DECLARATION

I hereby declare that no portion of the work referred to in this Project Thesis has been submitted in support of an application for another degree or qualification in this or any other university or institute of learning. If any act of plagiarism is found, I am fully responsible for every disciplinary action taken against me, depending upon the seriousness of the proven offence.

vi

# ACKNOWLEDGMENTS

I would like to thank my advisor, Professor Florin Rusu, who has equipped me with many technical skills in the Database field. He gives me chances to dive deep into database systems and code generation. During the current study, I really learned and gained much. Without his guidance, the study would not have been possible.

# ABSTRACT

**Code Generation Techniques for Raw Data Processing**

by

<u>Xin Zhang</u>

Master of Science

in

Electrical Engineering and Computer Science

University of California, Merced

2017

The motivation of the current study was to design an algorithm that can speed up the processing of a query. The important feature is generating code dynamically for a specific query.

We present the technique of code generation that is applied to query processing on a raw file. The idea was to customize a query program with a given query and generate a machine- and query-specific source code. The generated code is compiled by GCC, Clang or any other C/C++ compiler, and the compiled file is dynamically linked to the main program for further processing. Code generation reduces the cost of generalizing query processing. It also avoids the overhead of the conventional interpretation during achieve high performance.

Database Management Systems (DBMSs) perform excellent jobs in many aspects of big data, such as storage, indexing and analysis. Those amazing functionalities are implemented at the cost of versatility. DBMSs typically format entire data and load them into their storage layer. They increase data-to-query time, which is the cost time it takes to convert data into a specific schema and persist them in a disk. Ideally, DBMSs should adapt the input data and extract one/some of columns, not the entire data, that is/are associated with a given query. Therefore, the query engine on a raw file can reduce the cost of conventional general operators and avoid some unnecessary procedures, such as fully scanning, tokenizing and paring the whole data.

In the current study, we introduce our code-generation approach for in-situ processing of raw files, which is based on the template approach and the hype approach. The approach minimizes the data-to-query time and achieves a high performance for query processing. There are some benefits from our work: reducing branches and instructions, unrolling loops, eliminating unnecessary data type checks and optimizing the binary code with a compiler on a local machine.

**Keywords**: Code Generation, Query Processing, Template Approach, Raw File

# Chapter 1

# Introduction

## 1.1 DBMS

A database is an organized collection of data. It is the collection of schemas, tables, queries, etc. A database management system (DBMS) is computer software that interacts with clients or applications; it can also be used to analyze the data and mine information behind the data. A general-purpose DBMS is designed to allow users to define, create, query, update and delete data in databases. Well-known DBMSs include MySQL, PostgreSQL, MongoDB, Microsoft SQL Server, Oracle, and so forth. A database can be accessed through a Graph User Interface (GUI) or a Command Line in a terminal as well as through ODBC (Open Database Connectivity) or JDBC (Java Database Connectivity) to work with more than one DBMS. Database management systems are often classified according to the database model they support; the most popular database systems since the 1980s have all supported the relational model as represented by the SQL language.

There are many different kinds of database-manager systems. The in-memory database system uses the main memory, rather than a hard disk, as the storage layer. It is quicker than the other database systems that are built on disk. The state-of-art in a main-memory database system is MonetDB, whose internal data representation relies on the memory addressing ranges of modern CPUs. The cloud database is based on cloud technology. It can provide a storage service and a query-execution service to an endpoint client, such as a browser or an application with Open APIs. Nowadays, there are famous Cloud database vendors, such as the Amazon Relational Database Service, the Clustrix Database, the Microsoft Azure SQL Database and the Google Could SQL. They not only offer the services of relational database-manager systems but also provide services of No SQL Database systems, such as the Amazon DynamoDB, the Amazon SimpleDB, the Azure DocumentDB and the Google Cloud Bigtable/Datastore.

## 1.2   Query Engine

In this section, we present the concept and architecture of a query engine. There are four components of a Query Engine: First, the entry module of the Query Engine is the SQL parser, which is defined in the schema files of Flex and Bison. The SQL parser's function is to read the plain text of an SQL query and convert it into a parser tree. Each node in the parser tree represents each definition of a schema of the SQL parser. People can easily get each of the parser tree's nodes. For example, the TPC-H query 6 (modified) is only aimed at simplifying the problem and should be generated a parser tree like the following.

**Snippet 1.1**

```
SELECT
    l_discount
FROM
    lineitem
WHERE
l_shipdate >= date '1994-01-01';
```

```
[
  {
    "entityId": -101,
    "entityDt": "customerDt",
    "c_name": "Customer#000000001",
    "c_address": "IVhzIApeRb ot,c,E",
    "c_nationkey": -2015,
    "c_phone": "25-989-741-2988",
    "c_acctbal": 711.56,
    "c_mktsegment": "BUILDING",
    "c_comment": "to the even, regular platelets. regular, ironic epitaphs nag e"
  },
  {
    "entityId": -102,
    "entityDt": "customerDt",
    "c_name": "Customer#000000002",
    "c_address": "XSTf4,NCwDVaWNe6tEgvwfmRchLXak",
    "c_nationkey": -2013,
    "c_phone": "23-768-687-3665",
    "c_acctbal": 121.65,
    "c_mktsegment": "AUTOMOBILE",
    "c_comment": "l accounts. blithely ironic theodolites integrate boldly: caref"
  },
  {
    "entityId": -103,
    "entityDt": "customerDt",
    "c_name": "Customer#000000003",
    "c_address": "MG9kdTD2WBHm",
    "c_nationkey": -201,
    "c_phone": "11-719-748-3364",
    "c_acctbal": 7498.12,
    "c_mktsegment": "AUTOMOBILE",
    "c_comment": " deposits eat slyly ironic, even instructions. express foxes detect slyly. blithely even accounts abov"
  },
  {
    "entityId": -104,
    "entityDt": "customerDt",
    "c_name": "Customer#000000004",
    "c_address": "XxVSJsLAGtn",
    "c_nationkey": -204,
    "c_phone": "14-128-190-5944",
    "c_acctbal": 2866.83,
    "c_mktsegment": "MACHINERY",
    "c_comment": " requests. final, regular ideas sleep final accou"
  }
]
```

Figure 1.1 The structure of raw data

After the parsing, we can extract each expression of the parsing tree, such as the table of which we should take care, determination of the predicate expression, the column we should project and the kind of data we should filter.

In addition, the parser can check whether a query is valid based on the definition and schemas of Flex and Bison.

The next module we would like to mention is the query optimizer module.

To calculate the cost of a join order, we have some criteria for cost computing. We should consider five kinds of cost: 1, access cost to secondary storage, 2. storage cost, 3. computation cost, 4. memory uses cost and 5. communication cost.

In the five costs, the most expensive one is the cost of access to secondary storage.

The storage cost depends on the size and type of applications run on the database.

To estimate the cost of various execution strategies, we must keep track of any information that is needed for the cost function. For example, how many tuples it has and how many distinct values it has. This information is stored in a catalog, in which it is accessed by the query optimizer.

The most important time-consuming part is the Join order. It is because the final process of Join operations is to compute two or more tables and keep all the valid tuples together. This process involves much computation and memory. So, trying to keep this cost of the process as low as possible is crucial for the whole query process. Estimating the number of tuples after the JOIN operation requires the development of reasonably accurate cost functions for JOIN operations.

The JOIN operations define the relation containing tuples that satisfy a specific predicate F from the Cartesian product of two relations R and S.

**Different strategies for Join Cost Estimation.**

Table 1.1 The join estimate formulars

| Strategies | Cost Estimation |
|---|---|
| Block Nested-loop Join | A) numOfTuples(R) + (numOfTuples(R) * numOfTuples(S)), <br><br> if the buffer size is one tuple <br><br> B) numOfTuples(R) + [(numOfTuples(S) * (numOfTuples(R) / (sizeOfBuffer - 2))], <br><br> if a buffer for R has a buffer size of -2 tuples <br><br> C) numOfTuples(R) + numOfTuples(S), <br><br> if all tuples of R can be fit into a whole memory |
| Sort-Merge Join | A) numOfTuples(R) * [(log(numOfTuples(R))] + numOfTuples(S) * [log(numOfTuples(R))] <br><br> B) numOfTuples(R) + numOfTuples(S) |
| Hash Join | A) 3(numOfTuples(R) + numOfTuples(S)), <br><br> if the Hash index is in the meomory <br> B) 2(numOfTuples(R) + numOfTuples(S)) * [log(numOfTuples(S)) - 1] + numOfTuples(R) + numOfTuples(S) |

In TPC-H, for example, R is Customers and S is Orders. Estimating for Customers * Orders without using information about foreign keys: V(CName, Customers) = 5,000 and V(CName, Orders) = 2,500. Thus, the two estimates are 5,000 * 10,000 / 2,500 = 20,000 and 5,000 * 10,000 / 5,000 = 10,000.

We choose the lower estimate, which, in this case, is the same as our earlier computation using foreign key information.

The query optimizer module has two parts: one is to build the join tree, and the other is to build the topological chain, including the join tree we already generated.

For the join tree, there are 2 algorithms for the optimal evaluation plan: the first one is left hand join, which uses a greedy algorithm, and the second one is permutation join, which is a dynamical programming algorithm. We choose the latter one. The dynamical programming algorithm goes over each permutation of join order to compute how much it costs; thus, the dynamical programming algorithm is likely to give us the most optimal solution. Thus, we can gain the quickest join execution performance by the most optimal join order tree in the query execution stage.

After generating the query execution plan, the next stage is to generate the code for this particular query dynamically. In the template approach, we use information for the catalog to fill out the template, such as avoiding the data type check, reducing branches and unrolling a for-loop.

Then, we compile the generated code file to a shared object library and pass this binary-version file to the execution engine. The execution engine then loads the shared object to the program and executes it based on a query execution plan.

## 1.3 Query Strategy

In this section, we discuss the significant query strategy in cost-based query optimization, which is how to select an optimal join order for the multiply join query. Why the query strategy considers the join order is that the join part is be the most time-consuming part among all operations, such as a project, a scan and a selection.

There are also two strategies for generating the order point: one is the left deep join strategy, and the other is the permutation join strategy.

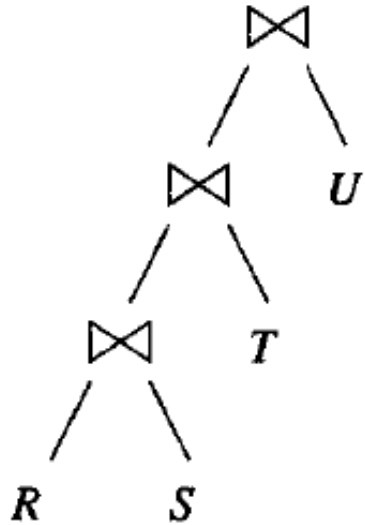

Figure 1.2 The join operation tree

In practice, the strategy of the left deep join tree is used very often. Figure 1.2 is an example of the left deep join tree. There are four advantages of using the left deep tree as a join order.

The total number of potential approaches to the shape of the left deep tree is large; however, because of all the possible shapes a tree with a given number of leaves can have, the number of possible left deep trees is not large. Therefore, searching for a better optimal query strategy would be quicker if we only consider the shape of the join tree is a left deep tree.

The left deep join tree is compatible with many common join algorithms, such as hash joins, nested loop joins and any other one-pass join algorithm.

If one pass algorithm is applied, the build relation on the left of each join only needs a smaller amount of memory.

If the nested loop algorithm is used, we can avoid generating any intermediate table more than once.

For the permutation join strategy, we go over each of combinations of the join order using dynamic programming. Dynamic programming is a method for solving a complex problem by breaking it down into a collection of simpler sub-problems, solving each of those sub-problems only once and storing its solutions for reducing latter computations. Dynamic programming algorithms are often applied for optimization of join orders. A dynamic programming algorithm examines the previous solution of sub-problems and combines the sub-problems' solutions to give the best solution for a given problem.

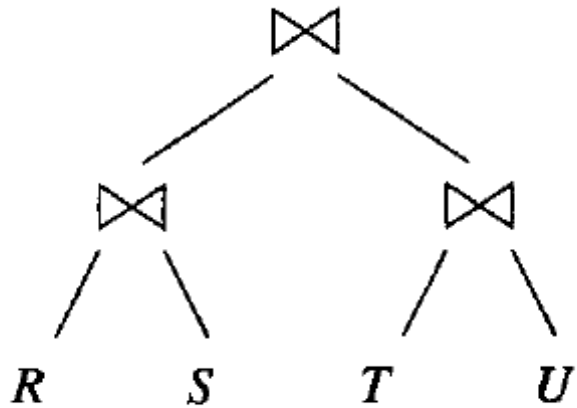

Figure 1.3 A join operation tree with deep balance

In a more practical manner, we use the dynamic programming algorithm as a table of cost and build the table with the cost of each sub-problem (subjoin), which we remember and put the intermedia result in an array and calculate using the previous result.

The shape of the most optimal solution by using dynamic programming can be a bushy one (Figure 1.3) or the left deep shape (Figure 1.2), depending on which one is the cheapest during the computation.

# Chapter 2

# Code Generation

## 2.1 Query Processing without Code Generation

When a query comes in, the parser checks whether it is a valid and supported query using Catalog/Metadata and a dictionary of token definitions of SQL. If it is well formed, then the query is passed to the Optimizer part; otherwise, it is returned immediately to the client with an error generated by the parser.

The optimizer builds a join tree and a topological execution chain. Based on the query text, all the join pairs are extracted with predicates. For the join tree, the optimizer uses the permutation algorithm to compute each possible join tree and its cost. The cheapest cost solution is chosen by the dynamical programming algorithm. Then, the join tree is built and ready to be put into the topological execution chain. After building the join tree of the query, other operations beside joins are considered for addition to the execution chain. The join operators of all the join nodes in the tree are all the same and have a general method. For example, each join operator checks every tuple, which is pulled from another operator, with its corresponding schema to extract the attributes of the tuple. Thus, we have to dynamically check each attribute's type (i.e. to indicate the codes from the project).

For example, the following figure indicates that the general method checks each attribute with all possible data types, such as Integer, Float as well as String and converts each tuple to the object with a specific schema.

```cpp
// then we convert the data to the correct binary representation
if (atts[i].type == Integer) {
    *((int *) &(recSpace[currentPosInRec])) = atoi (space);
    currentPosInRec += sizeof (int);
}
else if (atts[i].type == Float) {
    *((double *) &(recSpace[currentPosInRec])) = atof (space);
    currentPosInRec += sizeof (double);
}
else if (atts[i].type == String) {
    if (len % sizeof (int) != 0) {
        len += sizeof (int) - (len % sizeof (int));
    }
    string tmp(space);
    strcpy (&(recSpace[currentPosInRec]), space);
    currentPosInRec += len;
}
```

Figure 2.1 Data Type Checks in the General Approach

For scan operations, in the general approach without code generation, the scan operation loads data from a disk into memory by page unit, dynamically retrieves schemas for the data, and then translates each tuple of the page to an object/record with the schemas.

## 2.2  Query Processing with Code Generation

In contrast to query processing without code generation, after parsing the query, a specific code can be generated because the table name, schema, attributes' data type, etc. are already known. In the template approach, the information of a table and a schema can be filled into templates. For example, using template script language, such as M4, we can reduce the branches and replace them by unrolling the for-loop for each attribute and column. Thus, the generated file has limited number of branches and variables, which can decrease instructions of the code.

```
for each column from the column list
    curr=next;
    while ( (next-buffer) < MAX_LINE_SIZE && *next != 'SEPARATOR' &&
    *next!='\n') next++;
    *next = 0;
    next++;
    //column filling code
    ...
```

Figure 2.2. A Code Approach with Templates

For example, Figure 2.2 illustrates how to read each tuple from the data file. As the figure indicates, the for-loop has been used to go over each column, compute the offset and move to another column. The instructions that can be generated by the code have a number of jump and branch of instructions due to the for-loop, which can also affect the runtime efficiency in a CPU. Modern CPUs have superscalar features and a cache. Too many jump and branch instructions lead to miss hits and hard to take advantage of data locality.

```
                                    ;'
dnl # M4 CODE
dnl # declaring all the columns
m4_foreach(</_C_/>, </COLUMN_TYPES/>, </dnl
    // tokenizing
    curr=next;
    while ( (next-buffer) < MAX_LINE_SIZE && *next != 'SEPARATOR' &&
            *next!='\n') next++;
    *next = 0;
    next++;

    // column filling code
    M4_COL_FROM_TOKEN_ARRAY(_C_, i, curr)dnl
    M4_COL_MINMAX_ARRAY(_C_, i)dnl
/>)
dnl # END OF M4 CODE
```

Figure 2.3. The Template Approach

However, Figure 2.3 illustrates the template approach that incorporates GNU M4, which is an implementation of the traditional Unix macro-processor. The figure describes the unrolling of a for-loop for each column. Loop unrolling is used to reduce the number of jump and branch instructions that can potentially make the loop faster but increase the size of the binary. This gives an out-of-order or super-scalar CPU the possibility to schedule instructions better and thus run faster.

## 2.3 The Template Approach

In this section, we discuss code generation through the template-based approach. The basic idea behind the template approach is to use code templates for assorted queries, and fill out templates according to schemas and construct a single source-code file that evaluates the query dynamically.

The template-based approach can eliminate the defects which are brought by implementing general query evaluators in high-level abstract interfaces in most of modern database-manager systems.

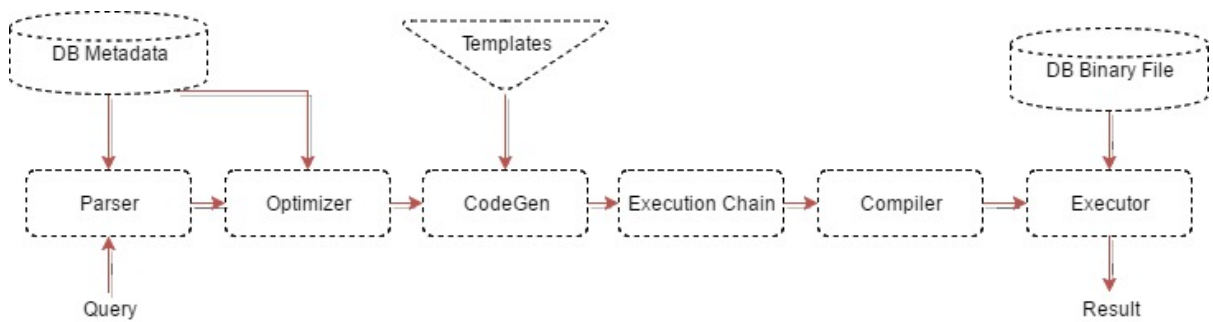

Figure 2.4 The overview of code generation

Figure 2.4 is the system overview of code generation through the template-based approach. Additionally, the generated code can employ the feature of the hardware, such as a CPU cache and pipeline.

The SQL parser first analyzes whether the query is valid according to the definition of schemas and catalogs. The next step is query optimization. Either the left deep tree algorithm with the greedy or the permutation algorithm with dynamic programming can be exploited to compute the cost of join orders and to choose the cheapest one at end of day. The output of the query optimizer forms a topologic chain with the join tree, in which each node includes a physical operator. The physical operators in the topologic chain should contain a scan operator, a select operator, a join operator, an aggregation operator, etc. Each operator is generated with hard-code functions in terms of information from the catalog.

By using M4 or any other technique, the code with hard-code methods is output to a single C++ source-code file. The compiler compiles the code into shared object library file. This binary file is then loaded to the main program, which executes the given query according to the previous plan. The output of the query will write back to the databases or build a view of the output.

The whole process of the code generation module is to traverse the chain, which includes a scan/selection, a join operator and an aggregate operator, and generate each codegen file for each operator. Finally, the codegen files are merged to one single codegen file.

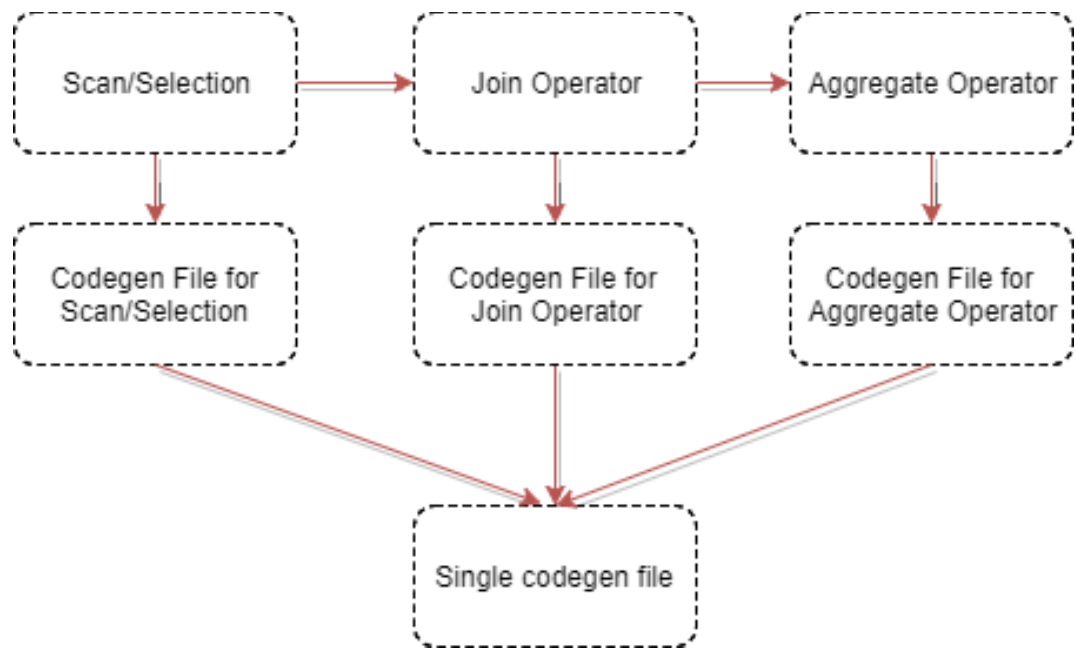

Figure 2.5 The process of code generation in operators

**Algorithm 1. The template of the join algorithm**

**Step 1 (Initialization)**

Declare a query bitmap that illustrates queries inside this chunk and other variables.

**Step 2 (Extract the left/right side columns from the input chunk and data-type check)**

m4_set_foreach(one column: the set of left/right columns)
m4_check (one column)
        m4_column_type(one column) * m4_column_data(one column) = NULL;
End for each

**Step 3 (Set up output columns and initialize spaces)**

m4_set_foreach(one column: the set of left/right columns)
   m4_check(one column)
   m4_column_type(one column) * m4_column_output_data(one column) = NULL;
End for each
m4_set_foreach(one column: the set of left/right columns)
   m4_column_out_data(one column) = (m4_column_type(one column)) *
   malloc(chunk_size * sizeof(m4_columns_type(one column));
End for each
………….

**Step 4 (Construct the join for each tuple)**

**Step 5 (Hash join)**

…………

**Step 6 (Clean up)**

m4_set_foreach(one column: the set of left/right columns)
m4_check(one column)
   free (m4_column_tuple(one column));
End for each

The template of join uses M4 to generate the code. In the template, m4_set_foreach means constructing the code by the set of left/right columns, and in each iteration, the iterator column is filled into the blank inside the foreach loop. The m4_check indicates the data-type check; in other words, in a general situation, the type of each column is determined by an if-condition dynamically. For example, a snippet of the code is as follows:

**Snippet 2.1**

```
Foreach(columns)
   if(type of column == String)
      then print(the index of columns + " type: String");
   else if(type of column == Integer)
      then print(the index of columns + " type: Integer");
   else if(type of column == Double)
      then print(the index of columns + " type: Double");
End for each
```

If we use the m4_check with the m4_foreach loop, supposing we have three columns, their types are separately String, Integer and Double; the code should be generated as follows:

**Snippet 2.2**

```
print(1 + " type: String");
print(2 + " type: Integer");
print(3 + " type: Double");
```

As can be n, the if-else branches are eliminated, and the generated code is so concise and short, which is only a three-line source code. Therefore, the code is executed quicker than before, since the if-else condition can significantly cause jump instructions and a cache miss, which happen inside the CPU.

Moreover, the foreach loop can also be erased. The foreach loop can lead to an unpredicted code path and jump instructions.

In the template, for the initialization part, each column can be assigned a fixed length of memory without using for loop and the data-type check. For instance, in step 3, we generate the code for each column with the column's type and chunk size; therefore, the hard code will be executed instead of the general code.

For the clean-up stage, the spaces each column has used are deleted and written in a hard-code manner. For example, step 6 should be as follows:

**Snippet 2.3**

```
Free(column1);
Free(column2);
Free(column3);
```

Code without code generation:

**Snippet 2.4**

```
Foreach(columns)
        Free(one column);
End for each
```

As the snippet of the code indicates, the for loop is eliminated by code generation, and the jump instructions are erased; thus, the code will run faster than the general one.

## 2.4 Bees Approach

In this section, we introduce one of the micro-specializations called Bees. This approach takes advantage of knowledge that variables have constant or fixed values during the runtime for a particular query. For example, an invariant comes from a schema and a catalog in a specific query context. That invariant is repeatedly employed in the hot execution path; therefore, it can significantly reduce needless operations as well as instructions and keep the code less and quicker, i.e. accelerate the speed of query processing for this specific input. But the values of invariants are known only at runtime, and the compiler cannot have optimized them during the compilation. So, the technique of code generation can be exploited.

**Snippet 2.5**

**Step 1. Read the input query.**

Input = getQuery();

**Step 2. Optimize the query and generate the plan.**

Plan = getQueryOptimization();

**Step 3. Execute the query plan**

Output = QueryExecution(plan, tables, schemas);

The above code snippet is the general query-processing algorithm. We read the query as the input, and the query plan is constructed by the query optimizer using information on catalog and schemas.

The query plan is executed by the query engine and will then write out the result as the output.

For the query optimization part, as mentioned in the previous chapter, the key to the optimization of a query is to make a decision that the join order is the cheapest and choose it as a query plan. This plan is actually a topological chain whose main part is the join tree.

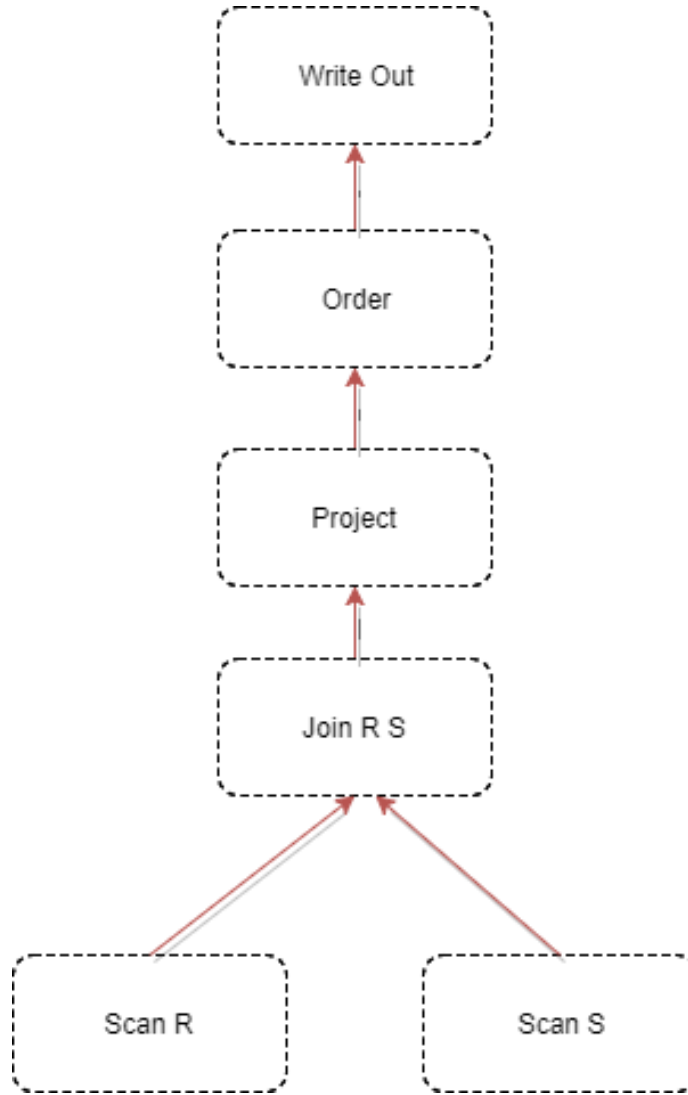

Figure 2.6 Query plan tree

Figure 2.6 illustrates a topological chain with the write-out node to the project node, which is the first segment of topological chain, and the Join node, which is the root of the join tree at the bottom of the topological chain.

In the volcano mode with tuple at a time, for example, the tuple iterates over each node/operator at a time. In each node, the tuples are parsed, extracted and cast to a data type by checking statements and if conditions.

The most time-consuming situation is when each tuple is checked repeatedly in each node for each query. For example, the schema is known, and the type and offset of each column are fixed. Thus, we do not need to excerpt the value of each column of each tuple; instead, we use invariants to replace the variables that are of an unknown data type and offset during the runtime because the schema and data type of columns' information are not known when the code is compiled.

In a database, many variables can be constant during the processing of a specific query. For example, the number of columns of a table and the length of a data type in each column are fixed when it is defined in the catalog. Generally, each tuple is stored in a sequential manner; if each tuple is processed when we know the offsets of each column, then it will help us to split the sequence of bytes easily without bothering the catalog during the runtime dynamically.

Moreover, we employ table *orders* from the TPC-H benchmark to demonstrate how bees can reduce the number of instructions.

**Snippet 2.5**

```
for(int i = 0; i < columns.len; ++i) //unrolling foreach
    if(columns[i] == null)
            then ...
    else ...
    value[i] = getColumnValue(i, size of datatype)

End foreach
```

In snippet 2.5, there are three kinds of invariant of which we can take advantage. The first invariant is according to the schema of *orders* in TPC-H. The value null is not available in the table *orders*; therefore, the checking condition for the value null in each iteration can be eliminated. In addition, any schema whose column value is not null will be optimized by bees. The second constant for any specific table is the number of columns; for example, it is 9 in the schema of orders. Then, we can unroll the for loop to reduce the jump instructions and incremental variable in the for-loop statement. The third constant is the offset of each column. For instance, the first column of the relation of *orders* is a 4-byte integer. Additionally, in each iteration, we do not to use the getColumnValue function to obtain the specific column value. Instead, we assign the value to the array with an index directly by using the invariant offset, which is 4 bytes for the first column in this example. Because the second column of the table *orders* is still an integer, the second offset is 4 bytes.

After the code generation of bees, code Snippet 2.6 is formed as follows:

**Snippet 2.6**

```
value[0] = *(int*) column;
value[1] = *(int*) (column + 4);
value[2] = *(Boolean*) (column + 8);
value[3] = *(double*) (column + 9);
…
```

After bees' code generation, as indicated in snippet 2.6, the code only has a few lines, specifically 9 lines, which equal the number of columns of table *orders*. Note that the column variable is an array of bytes, in which the sequence of data in one tuple is stored.

The main idea of micro-specialization, i.e. bees, is to analyze variables whose runtime value is constant for a specific query. They can replace the variables, return the value of a function or a block of code with the invariant value. For example, compare snippet 2.5 with snippet 2.6. The invariant value instead of the return value of the getColumnValue function.

There are many kinds of micro-specialization (bees); in the above example, they are relation bees. Each table has its own bee in DBMS. Since the code is replaced by invariant value and is typically very short and powerful for this specific query, its features resemble distinctive bees.

In addition, the term "bees" is a somewhat abstract concept; we call each particular routine of micro-specialization bees when we apply one or more invariants of some code block in the source code. For example, we use micro-specialization to replace the getColumnValue function; this can be called bee routine GCV, which is an acronym for getColumnValue.

As snippet 2.6 illustrates, different parts of the process, such as schemas and plans when relations are defined in a DBMS, can be invariant. When a query comes in, the table list, predicates and select list are fixed after query compilation. All of these can be advantageous. Thus, there are three kinds of bees.

The relation bees we describe above can use an invariant to replace the checking condition of a data type, an offset variable, the number of columns for unrolling a loop, etc.

Additionally, another micro-specialization is called query bees. In query bee, the variables for processing a query are fixed when the query is received and known in the database. For example, the query may include some predicated conditions, which can be "l_shipdate = '1990-01-01'". Then, the predicate statement consists of the column's name, which is l_shipdate; the operator, which is "="; and the constant value for this column, which is "1990-01-01".

The concept of bees can be developed further to the tuple level. The main idea is to put a set of possible values of the column into the code if the column has only a few values. Thus, the length, the offset and the data type of the column are fixed. So, this approach is called tuple bees. For example, the column "level" only has three values, the first is "high", the second is "mid", and the third is "low". The specific tuple bee of each column in this tuple is presented as a small index, which is from 1 to 3; it is termed beeID. Then, all the tuples that include the column can use beeID instead of the previous value. So, in this case, there are two tuple bees that are used for all the tuples in this table.

Another important scenario is when the bees generate the code. It should happen when the bees receive information they require. At the same time, the length, the offset, the data type of each column is fixed and constant. However, the instantiation time of each kind of bee is different; for example, relation bees are instantiated when the schema is created or alerted. Query bees are initialized and invoked when the query plan is constructed. The predicate of a query will be fixed, and query bees can optimize it during the runtime. Tuple bees work whenever a tuple is inserted into a table or updated. Additionally, these three kinds of bees can be initialized during compilation.

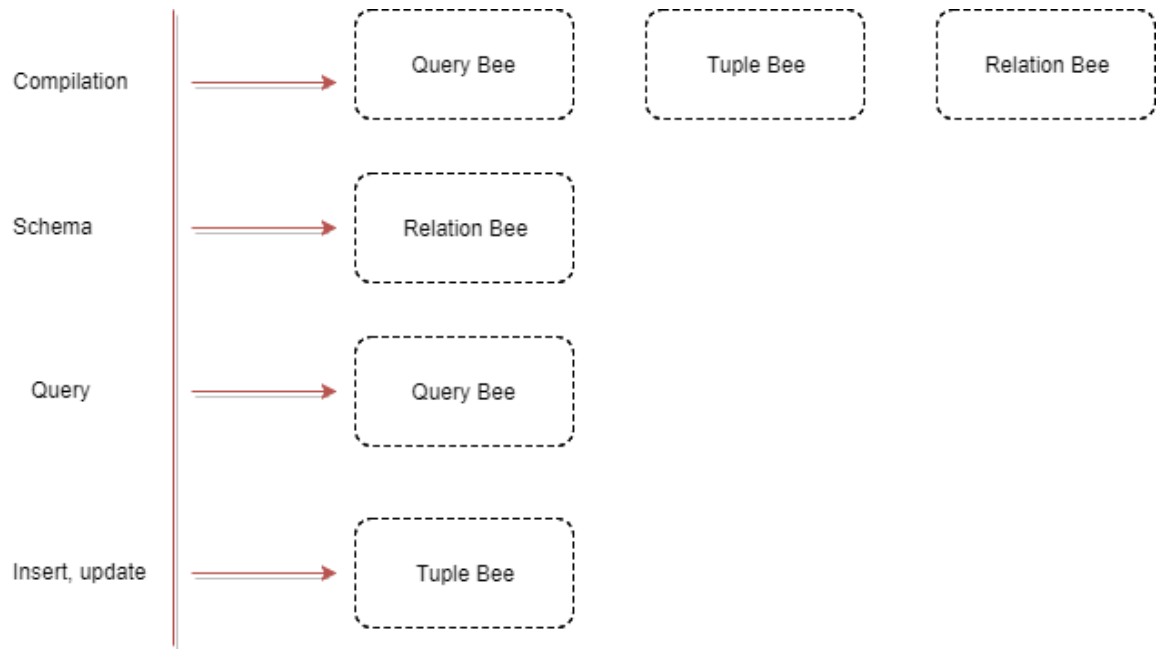

Figure 2.7 Query Bee tiers

There are two strategies for implementing bees in detail. One is the templates with variable blanks can be used, during the runtime, when the values of the invariant are known and then filled out. This method is exactly the template approach we discussed earlier.

Another strategy is when the invariant value is employed in the if-condition statement and has a few available values it could be; therefore, the if-condition or checking condition can be eliminated. As illustrated in snippets 2.5 and 2.6, all the data-type-checking conditions and all the for loops are removed.

Let us look at Impala, since it implements code generation using bees techniques.

Apache Impala is an open-source, native analytic database for Apache Hadoop. It is a massive parallel processing SQL query engine. The Impala project was launched in 2008. Impala was integrated with the Hadoop ecosystem; it can process a query over Hadoop File System (HDFS) files and any other Hadoop storage, such as HBase. In addition, it supports different file formats, such as text, Avro and Parquet.

For the code-generation part, it uses Just-In-Time(JIT) to compile a query. The tool it uses to implement JIT is a Low Level Virtual Machine (LLVM).

The LLVM is a compilation infrastructure that is a set of a reusable compiler and tool-chain technologies. It can help in developing front ends or back ends of a compiler. It is written in C++ language and designed for optimizing a program that is written in any other programming language.

The LLVM project started in 2000 at the University of Illinois in Urbana-Champaign. It was used only as a research framework to investigate dynamic compilation techniques for static and dynamic programming.

Impala uses code generation to eliminate three kinds of situations. The first kind is removing checking conditions. As we discussed earlier, the if-conditions can cause many jump instructions, which can significantly lead to a cache miss inside a CPU. However, with code generation, if-statements can be transformed to single cases; therefore, it reduces the jump instructions and take advantage of features of a CPU, such as the pipeline and single instruction multiple data (SIMD), which is a class of parallel computers in Flynn's taxonomy. Code generation techniques can allow a CPU to simultaneously compute multiple data based on a single instruction. The second Impala did is removing the load part. Loading data from memory is a pricey and CPU-unfriendly operation, which can lead to a cache miss and break the CPU pipeline. If the data loaded from memory are always the same copy of data, Impala can use code generation to replace this functionality of code to avoid loading the same data repeatedly. The third is removing virtual function calls. The virtual function calls relate to scan information of a function reference inside of a virtual table, which is called a vtable; this is a lookup table of functions that helps a program to invoke a correct function in a late-binding manner dynamically. Using the vtable can be very pricey. It causes many penalties. However, if the type of function is fixed during the runtime, the virtual function call can be replaced with the authentic function. In Impala, an expression is represented as a tree that is implemented by overriding a virtual function. For example, a customer writes a User Define Function (UDF), which is used to calculate the sum of two integers. Its expression is a tree, as illustrated in Figure 2.8.

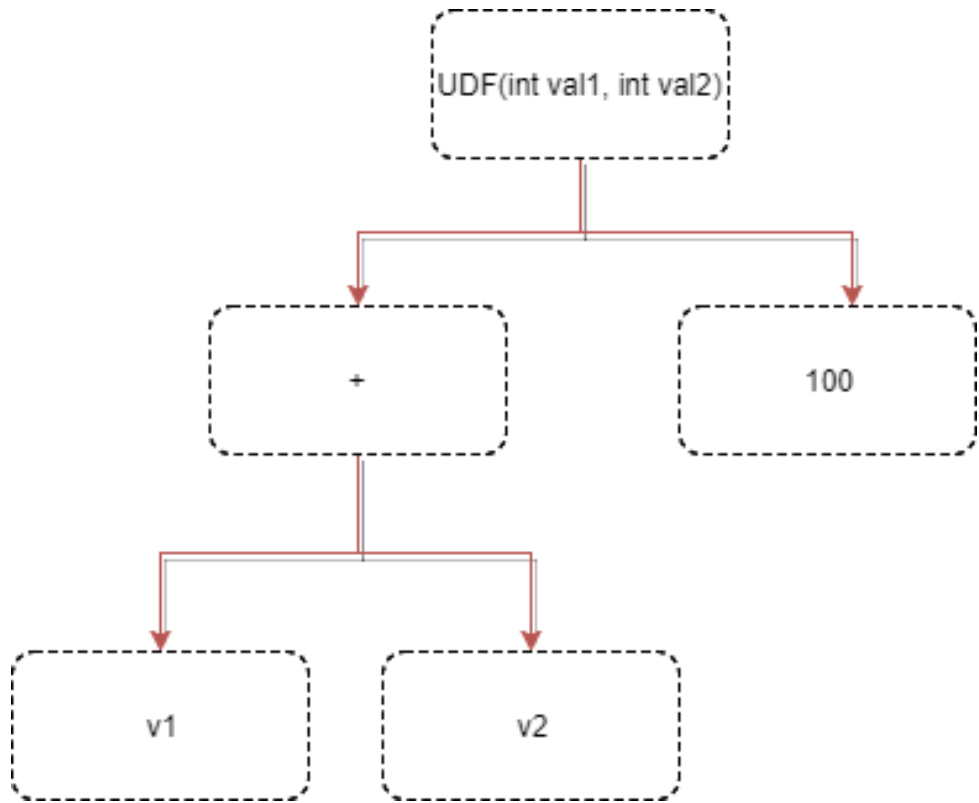

Figure 2.8 The UDF execution tree

The definition of UDF is demonstrated in snippet 2.7.

**Snippet 2.7**

```
Expression UDF(Expression v1, Expression v2) {
        Return (v1 + v2) * 100;
}
```

In the above snippet code, each expression is a virtual function; thus, the UDF invokes at least three virtual function calls. Impala resolves this issue by generating a code with inlining of the resulting function calls. Thus, there is no virtual function call any more. In addition, inline functions can increase the chance that the program runs parallelism inside the CPU and takes advantage of its features, such as the pipeline and SIMD.

### 2.4.1 Details of Impala techniques

Impala employs LLVM to generate and compile a code for a particular query in the backend module. There are three techniques that LLVM Impala uses: LLVM intermediate representation (LLVM IR), IRBuilder and Compilation to IR.

**LLVM IR**

The main concept of LLVM is IR, which resembles assembly language. Each machine instruction can be represented by each IR code. The Clang, which is the frontend of LLVM, can generate IR and be applied for some optimizations in the generated machine code.

**LLVM IRBuilder**

LLVM contains a series of tools; one of them is IRBuilder, which provides C++ API interfaces. Intermediate representation can be generated using IRBuilder. The IRBuilder code is transferred to IR through one instruction by one instruction. However, it is annoying that the size of the code generated by IRBuilder is larger than the C++ code with the same functionality. It makes the code hard to read after its generation.

Compilation to IR

Impala commonly uses Clang to compile a function written in C++ to IR and then inject this snippet code into an associated function during runtime. In Impala, most of code generation uses C++ functions that are compiled into IR by Clang; it is easy for developers to debug the logic of the code.

The intention of compiling the function into IR is to replace virtual function calls. When the feature of code generation is on, Impala recursively looks up all virtual function calls to expressions and replaces them with calls to code generation functions.

In Impala, the code generation cannot happen through only using IRBuilder. However, Impala is developing an infrastructure to edit generated IR instructions. It can help impala to unroll the for loop and avoid loading the same data repeatedly.

**Snippet 2.8**

```
SELECT
    sum(l_extendedprice * l_discount) as revenue
FROM
    lineitem
WHERE
    l_shipdate >= date '1994-01-01'
    AND l_shipdate < date '1994-01-01' + interval '1' year
    AND l_discount between 0.06 - 0.01 AND 0.06 + 0.01
    AND l_quantity < 24;
```

Snippet 2.8 is query 6 of TPC-H. We run it over an Impala system with a log mode. Snippet 2.9 is the log of the stack of processing query 6 with code generation. First, from the stack log Impala provided, we can notice that the query was received by QueryExecMgr. Therefore, the execution plan was generated by PlanFragmentExecutor. After the execution plan was built, Impala instantiated the aggregation node with code generation.

**Snippet 2.9**

```
   2049 I0409 17:01:53.064020 18846 status.cc:114] Timestamp not yet
supported for codegen.
   2050@0x7f3f85b8584d  impala::Status::Status()
   2051@0x7f3f84ab4ce9
impala::HdfsScanner::CodegenWriteCompleteTuple()
   2052@0x7f3f84af31e6  impala::HdfsTextScanner::Codegen()
   2053@0x7f3f84a94888  impala::HdfsScanNodeBase::Codegen()
   2054@0x7f3f84a43baa  impala::ExecNode::Codegen()
   2055@0x7f3f84bb03c5  impala::PartitionedAggregationNode::Codegen()
   2056@0x7f3f8530dea3  impala::PlanFragmentExecutor::PrepareInternal()
   2057@0x7f3f8530c07d  impala::PlanFragmentExecutor::Prepare()
   2058@0x7f3f852ec7c5  impala::FragmentInstanceState::Exec()
   2059@0x7f3f8531627b  impala::QueryExecMgr::ExecFInstance()
   2060@0x7f3f853190b2  boost::_mfi::mf1<>::operator()()
   2061@0x7f3f85318f3b  boost::_bi::list2<>::operator()<>()
   2062@0x7f3f85318a6f  boost::_bi::bind_t<>::operator()()
   2063@0x7f3f8531867a
boost::detail::function::void_function_obj_invoker0<>::invoke()
   2064@0x7f3f857c77f4  boost::function0<>::operator()()
   2065@0x7f3f857c4e87  impala::Thread::SuperviseThread()
   2066@0x7f3f857ce380  boost::_bi::list4<>::operator()<>()
   2067@0x7f3f857ce2c3  boost::_bi::bind_t<>::operator()()
   2068@0x7f3f857ce286  boost::detail::thread_data<>::run()
   2069@0x87fa2a  thread_proxy
```

Snippet 2.10 is the source code of PartitionedAggregationNode, which is in charge of calculating the aggregation operation; it directly invokes the ExecNode::Codegen(state) to generate code recursively. ExecNode is the supper class of all executor nodes. Since it is supper class, the subclass that will be called is unknown when the code is compiled. The subclass would be invoked correctly only in runtime. Another important reason is that the type of parameter of ExecNode::Codegen(state) is the RuntimeState class, which is a collection of all items in global state of a query that is shared across all execution nodes of that query.

**Snippet 2.10**

```
void PartitionedAggregationNode::Codegen(RuntimeState* state) {

    DCHECK(state->ShouldCodegen());
    ExecNode::Codegen(state);
    if (IsNodeCodegenDisabled()) return;

    LlvmCodeGen* codegen = state->codegen();
    DCHECK(codegen != NULL);
    TPrefetchMode::type         prefetch_mode         =         state_-
>query_options().prefetch_mode;
    Status codegen_status = is_streaming_preagg_ ?
        CodegenProcessBatchStreaming(codegen, prefetch_mode) :
        CodegenProcessBatch(codegen, prefetch_mode);
    runtime_profile()->AddCodegenMsg(codegen_status.ok(),
codegen_status);
  }
```

Since query 6 has a predicate for selection, HdfsScanNodeBase::Codegen() is employed. HdfsScanNodeBase is the base class for all HDFS scan nodes. There is a prerequisite that we have loaded all the TPC-H data with factor 1 into the HDFS. We use an Impala shell to load the TPC-H data we generated using DBgen tools on disk to the HDFS using another technique, the keyword called LOCATION, after defining schemas of TPC-H. The load can allow us to put the data into HDFS easily, as illustrated in snippet 10.

Since HdfsScanNodeBase is a base class, the real class should be HdfsTextScanner::Codegen() in the runtime, which is an implementation of the supper class of HdfsScanner that understands text-formatted records. One of techniques it uses is Streaming SIMD Extensions (SSE) instructions. SIMD instructions can greatly increase performance when exactly the same operations are to be performed on multiple data objects.

**Snippet 2.11**

```
CREATE EXTERNAL TABLE lineitem
        (
         l_orderkey INT,
            l_partkey INT,
            l_suppkey INT,
            l_linenumber INT,
            l_quantity INT,
            l_extendedprice FLOAT,
            l_discount FLOAT,
            l_tax FLOAT,
            l_returnflag STRING,
            l_linestatus STRING,
            l_shipdate TIMESTAMP,
            l_commitdate TIMESTAMP,
            l_receiptdate TIMESTAMP,
            l_shipinstruct STRING,
            l_shipmode STRING,
            l_comment STRING
         )
ROW FORMAT DELIMITED FIELDS TERMINATED BY'|'
LOCATION '/tpch/lineitem';
```

HdfsTextScanner::Codegen() eventually calls the CodegenWriteCompleteTuple() function to write out each tuple with evaluate conjunctions. It is fully implemented by LLVM IRBuilder, which is one instruction by one instruction to assembly language. So, it seems like people write a function using assembly language.

By contrast, if Impala does not use code generation for writing each tuple. The source code is short and concise without code generation, since in this case, the functionality is fully implemented by using high-level language rather than using assembly language such as LLVM IR. Finally, if everything works fine, the OK status should be return, as the stack, Snippet 2.9, indicates.

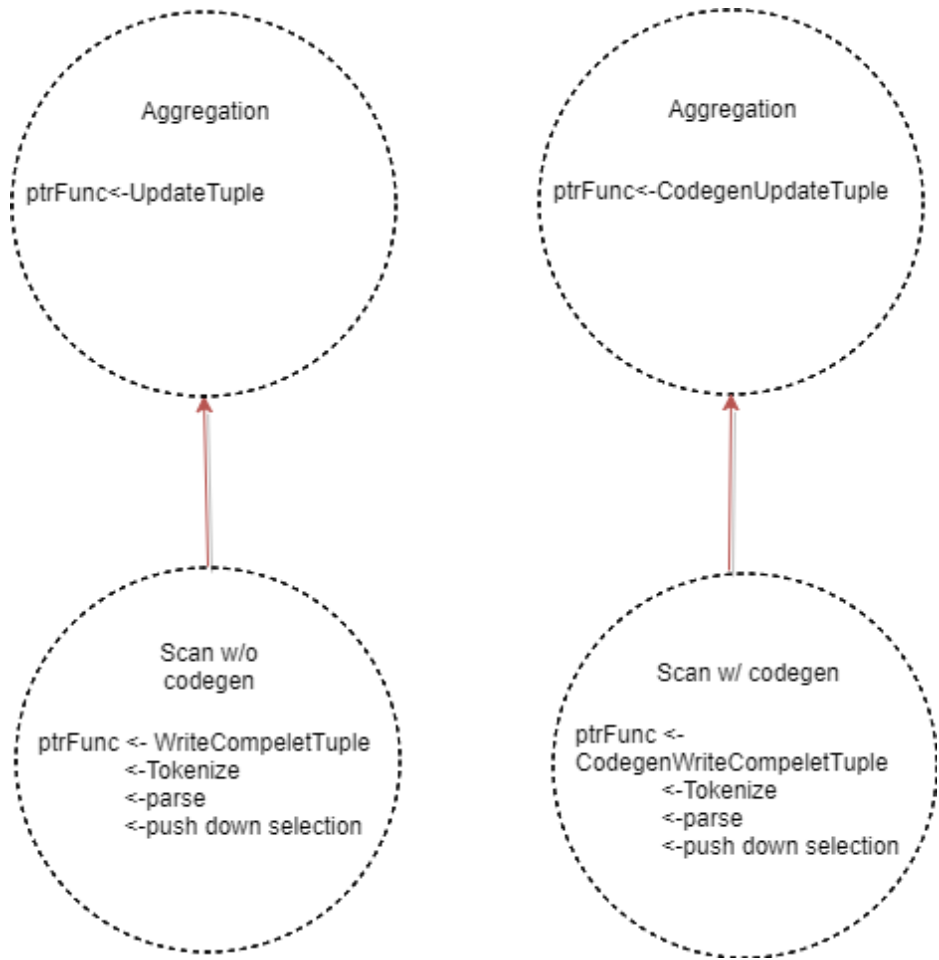

Figure 2.9 Code generation in Impala

Figure 2.9 illustrates two situations: the left two nodes represent processing query 6 without code generation, and the right two nodes represent processing with code generation.

## 2.5 HyPer Approach

In this section, we discuss another technique of code generation called HyPer. It gives people a brand-new idea of how to generate a code without the iterator model. Nowadays, the majority of database systems reads a query and translates it into a physical algebra expression, which is a topological chain with a join tree. This is a traditional method of building an execution plan, which is called the iterator model or volcano processing model. In this model, each algebraic operator node produces a tuple, and the input comes from lower level operator nodes by calling the next function recursively, which is always a virtual function call.

The advantage of the iterator model or tuple-at-a-time model is that it is easy to implement. However, its drawback is that it is not CPU friendly. The reasons are following:

First, the next function is typically a virtual function, and a virtual function call can incur the action of looking up the vtable. The vtable needs additional instruction to determine whether the runtime function is the correctness function. It can cause a branch misprediction, thereby breaking the pipeline inside the CPU since jump instruction causes many bubbles in the pipeline. Additionally, each virtual function call happens if and only if a tuple is produced in an operator node. Imaging that one tuple can trigger the number of operator nodes times virtual function call. Therefore, it is a significant penalty in terms of runtime tons of tuples processed.

Second, this model is not good for data locality. The core of the iterator model or tuple-at-a-time model is the topological chain, which includes algebraic operators. The iterator model is for operator centric, and data or tuples are pulled from one node to another node; in the iterator processing, it may involve data/tuple copy, which is against the pipeline feature. Note that the definition of the pipeline is that an operator can pass data to its parent operator without copying, otherwise materializing the data.

Right now, some of database systems, such as Impala, use the block-at-a-time model, which is pulled a bunch of tuples at a time rather than one tuple. This model can amortize the costs of virtual function calls. However, it still breaks the pipeline since the data or tuple copies commonly happen. At the same time, breaking the pipeline causes an increase in the usage of the main memory and a cache miss inside a CPU.

The point of view of HyPer processing is data centric and not operator centric. Data are processed that can keep tuples in CPU registers as long as possible. Data are not pulled by operators but pushed towards the operators. This results in much better data locality.

The query-processing architecture of HyPer

There are two terms we should explain before introducing the query-processing architecture. One is the pipeline breaker, which is a physical algebraic operator that puts a processing tuple out of register in the CPU. The other term is the full pipeline breaker, which copies all the tuples before processing them.

The intention of HyPer code generation is to maximize the data locality and achieve high performance of query processing.

HyPer assumes that the number of registers is large enough for columns in each tuple since HyPer takes advantage of the feature that an unlimited number of registers can be used in LLVM.

The traditional tuple-at-a-time model is not CPU friendly. Tuples are pulled by calling the next function, which can cause a cache miss as tuples can be out of registers. The block-at-a-time model, which is applied in Impala, passes more than one tuples at a time; it is typically a few chunks of tuples. However, it still breaks the pipeline since the number of tuples may beyond the bounds of the quantity in the registers in CPU. Actually, any tuple-at-a-time or block-at-a-time paradigm pulls data up from another operator and can break the pipeline during runtime.

HyPer changes the direction of data flow. Tuples are pushed from one operator to another rather than pulled, i.e. the so-called push model compared to the volcano model. Tuples in each node are computed in a more affordable manner since tuples are processed without a large number of virtual function calls. Once the tuples are done in a node, they are pushed to another node. In this push-based model, the control flow is out of the inner loop inside of each operator. In other words, in the previous iterator- or tuple-at-a-time model, the control flow is inside of each next function, since one tuple is always pulled through the leaves, which are typically scan operators, of the execution tree to its top operator. Therefore, the control flow cannot be interrupted from leaves to the root. In addition, the push model can significantly reduce the number of materializations of tuples.

**Snippet 2.12**

```
select *
from R1,R2,
where R1.x = 7 and R1.a = R2.b
```

**Snippet 2.13**

```
initialize memory of the hash table of join a = b
for each tuple t1 in R1
        if t1:x = 7
        materialize t in the hash table of Join a=b
for each tuple t2 in R2
        for each match t1 in Join a = b[t2:b]
output t1 × t2
```

Without the iterator model, in HyPer, the query-specific code should be snippet 2.13, the corresponding SQL query is illustrated in snippet 2.12. It selects relation R1 with a value of 7 in column x. and then R1 joins R2 together with the column a of R1 and the column b of R2. The code of snippet 2.13 is the exactly translated version of the query illustrated in snippet 2.12. As can be seen, the code of snippet 2.13 contains four parts. The first is initialization of the hash table for the join operation. The second part is scanning and selection, which filter with the tuples whose value is 7. The third fragment code is the join part. Each tuple inside relation R2 is joined with all tuples that are selected by the second fragment code. The last part of the code is outputting the result. All of the above four code sections are pipelining, meaning that HyPer can keep tuples in the register of a CPU when they are processed at runtime. As can be seen, this structure of code results in high localized data. Tuples are produced in the tight loop rather than passing each tuple through leaves to the root of the execution tree.

How to translate a query into code fragments.

We noticed that the boundaries of a query-specific code are blurred. Each code section produces all the tuples or some of them at a time. In other words, the generated code is a data-centric strategy, not an operator-centric strategy. In practice, each operator can contain multiple code fragments. Therefore, it is harder to translate a query into a query-specific code than generate a tuple-at-a-time code. In addition, each fragment code is not very structural, and it is too flexible to generate. In the tuple-at-a-time model, every operator only calls the next function, and the whole routine is controlled by the execution chain. By contrast, in HyPer, it is hard to express the code, especially if the logic is complex. However, doing code fusion is necessary since the generated code consists of all the code fragments that can keep all columns of each tuple inside CPU registers.

In HyPer, it offers an interface that is almost as simple as in the tuple-at-a-time model. Every operator provides two functions: Produce() and Consume(columns, source).

The Produce() function requests the operator to produce its result tuples. When the operator has finished, tuples are pushed back to the caller function through invoking their consume functions. For snippet 2.13, as an example, join.produce() calls the left-side select.produce() to filter the tuples from the result tuples of a scan. Then, scan.produce() is called, which scans the tuples in the relation on disk; it is also a leaf node in the tree, so the invoke chain stops going deep at this operation node. Upon finishing its job, the scan operator calls the select.consume() function back with the result tuples as parameters to push back to the select operator. Then the selection node filters the result tuples that come from scan operator and passes all the tuples whose value only satisfies the predicate statement. After all tuples are produced in selection nodes, the control flow goes back to the join node through the join.consume() call. Then, the join node invokes the select.produce function of the probe side of operation, which is typically the right side of the join part. Then, the right side does the same procedure the left side did previously. Once tuples from both sides are produced in the join node, the join algorithm can be applied to generate the final result tuples. However, in HyPer, the produce and consume interfaces are only an image model. They are used only by code generation.

When the query comes in, the query is first translated into a physical algebraic plan. The different part from the other system is the plan that is not translated into an executable topological chain; instead, it is compiled into the fragments' code. Only in this step, the interface produces and consume is used to generate the code.

Creating a Machine Code

The conventional approach to generating a code is to generate a C++ code from the query and compile it into a shared object by using a compiler such as G++. Then, the code is loaded into a program and run. However, there are some drawbacks. First, the compiler takes long time to compile the code; sometimes, it takes seconds. Second, once the compiler is used to compile the code dynamically, people typically use the shell command to do query, meaning that the control of data flow is taken by the execution of the command out of the program itself. However, in HyPer, LLVM is used to generate the IR code, which can be executed in any other machine and optimized by JIT. In addition, the LLVM JIT builds the assembly code that could be machine-specific code. The speed of this code's compilation (a few milliseconds) is higher that implemented in using GCC/G++ compiler.

Writing assembly code is difficult even when the LLVM IRBuilder is used. Therefore, HyPer used the mixed model to generate the code. The complex parts are implemented in C++, and the connect code between operators is written using LLVM IRBuilder. All the operators are pre-compiled; in other words, they are compiled into the LLVM IR code before running the code. The connection code is dynamically generated. So, the compilation time should be super-short.

**Snippet 2.14**

```
  define internal void @scanConsumer(%8— %executionState, %Fragment R2— %data) f
    body:
    ...
    %columnPtr = getelementptr inbounds %Fragment R2— %data, i32 0, i32 0
    %column = load i32— %columnPtr, align 8
    %columnPtr2 = getelementptr inbounds %Fragment R2— %data, i32 0, i32 1
    %column2 = load i32— %columnPtr2, align 8
    ... (loop over tuples , currently at %id, contains label %cont17)
    %yPtr = getelementptr i32— %column, i64 %id
    %y = load i32— %yPtr, align 4
    %cond = icmp eq i32 %y, 3
    br i1 %cond, label %then, label %cont17
    then:
    %zPtr = getelementptr i32— %column2, i64 %id
    %z = load i32— %zPtr, align 4
    %hash = urem i32 %z, %hashTableSize
    %hashSlot = getelementptr %"HashGroupify::Entry"— %hashTable, i32 %hash
    %hashIter = load %"HashGroupify::Entry"— %hashSlot, align 8
    %cond2 = icmp eq %"HashGroupify::Entry"— %hashIter, null
    br i1 %cond, label %loop20, label %else26
    ... (check if the group already exists , starts with label %loop20)
    ... (more loop logic)
  }
```

As snippet 2.14 illustrates, the complicated part of the scan is written in C++. It has been pre-compiled into LLVM IR instructions. in addition, the control flow is generated using C++. But for tuples, such as accessing tuples and assemble tuples, are all written in the LLVM IR assembly code. For higher performance, HyPer implements the hot path by LLVM IR.

Join Operators

The implementation of the scan operator and the selection operator is clear and simple. By contrast, the join operator is the more complicated part. The reason why it is complex is that more than one time transition of the control flow can happen. For example, the join algorithm can run a tuple operation with LLVM IR, so the control data flow is taken by LLVM just after the join operation is employed; then, the control flow can oscillate from LLVM to C++. In addition, the complicated join algorithm can make the generated code extremely large if and only if there are multiple join operations in the query. Therefore, it needs multiple functions for the join operator at runtime. The advantage of HyPer is to keep track of all columns and remember whether they are in the register. Materializing in memory is rather quick, but for implementation, it is very complicated to design.

In code snippet 2.14, the LLVM IR loads the pointers to the columns that are ready to be accessed. Then. It goes over each tuple and loads an associated column to a register and evaluates the column with the predicate statement. If the predicate statement is not satisfied with the condition, the loop continues. Otherwise, the LLVM IR loads another associated column to a register. Thus, the hot paths are always written in LLVM and may invoke a native C++ function if needed.

## 2.6 The Relaxed Operator Fusion Approach

In this section, we discuss another approach that is similar to HyPer in the last section; it is called the Relaxed Operator Fusion Approach. Modern CPUs can provide two significant features that can improve programs to run faster and more efficiently. One is the SIMD technique, which is a class of parallel computers in Flynn's taxonomy. It illustrates machines with multiple processing elements that simultaneously perform the same operation on multiple data. It is different from multithread, which is multiple instructions that results multiple data, which can bring some concurrent issues. The scenario for SIMD is to apply a single instruction to vectorize data. For example, it one would like to obtain all values less than 10, the naive approach is to go over each item and compare it with value 10. If it is less, then it is put into a collection or set; otherwise, it is discarded. Whether people use SIMD, however, depends on the size of the CPU cache. Suppose the cache size is 10 KB, each tuple is 1 KB, and the CPU can thus perform 10 tuples at a time. This is definitely faster than the tuple-at-a-time model.

The other technique is called perfecting. Its instructions can move chunks of tuples from DRAM to CPU registers before those chunks are used in the CPU. Thereby, the overlap of the time between the previous iteration and the next iteration is used efficiently. For example, when the CPU processes a bunch of tuples, the I/O is idle. The I/O can be exploited to load the tuples for the next iteration from the main memory to CPU caches; thus, when the next iteration is executed, its data are ready to be consumed in the CPU caches. That is great idea of hiding the latency of expensive cache misses.

The authors combine those two advanced techniques through code fusion to achieve high performance and data locality. People call this combination relaxed operator fusion (ROF). ROF can allow DBMS to generate the code with SIMD vectorization and the prefetching technique.

In terms of generating the code, it is similar to HyPer code fusion. First, the authors define the term "pipeline breaker". Data locality can only happen, without spilling tuples into disks, if all operators of one pipeline are kept in one for loop to maximize the data locality. However, the difference is that the HyPer paper uses the tuple-at-a-time model. In other words, only one tuple is processed in one moment, which is different from that of the ROF. By contrast, the ROF can take advantage of SIMD; with one tuple, the SIMD cannot achieve high performance. Therefore, ROF uses a vector at a time instead of a tuple at a time. In addition, this model can prefetch a bunch of tuples for the next iteration.

# Chapter 3

# Query Processing over Raw Data

## 3.1 Introduction

Raw data or a raw file is data collected from a source. The raw data are unprocessed, and they can be in a structured file format, such as CSV, XML and JSON. Alternatively, they can be unstructured data, such as a web application log. The raw data are more readable than binary data, and humans can thus extract information directly from raw file. Nowadays, as the time of the data flood has come and data collections become larger and larger, much data or stream is appended to files or inserted directly into a database every second. However, inserting each tuple or a piece of data into a database is very expensive, as it includes many procedures such as transactions, indexing and key constriction. Some solutions invented by a few database experts are loading the data directly from a raw file to a database. By using a database, people can take advantage of most powerful features inside the database. For example, indexing is commonly implemented by a B+ tree and can take log(N) time to search a key. However, the problem is that the raw files are typically large, and it takes a long time to load to a database. In addition, if we want to use indexing or any other functionality, it still needs to build or construct data structures in advance, which also takes time. In reality, in most scenarios, people read raw data only to do some analysis or select some columns once and then discard the data. This means that loading the raw file into a database is unnecessary and too expensive.

Then, the execution of a query over the raw file comes out. It can run queries (load, select, join, etc.) on the raw data without using any kind of database. Some databases support doing a query over a raw file, which is called the external table.

The external tables feature is a complement to existing SQL*Loader functionality. It enables people to access data in external sources as if it were in a table in the database. An external table is a table that is not inside a database but is transparent to users, which means users cannot be aware that the raw file is outside the database. Typically, people can send a query, which needs some information in the raw file, to a database. The performance is limited depends by the speed of the file scan, tokenizing and parsing.

## 3.2 SCANRAW

In this section, we introduce our novel physical operator of a database for in-situ processing, which combines data loading and an external table seamlessly. It has a parallel super-scalar pipeline architecture that can overlap the scanning raw file by converting tuples from raw data into the database format and processing for queries. Additionally, another mechanism we invented is *speculative loading,* which can format rows from raw data into database when there is any interval time frame, and the read buffer is full. Its motivation is when the CPU is busy processing a query, the I/O is idle. Therefore, we can take advantage of the I/O idle time to minimize the total processing time.

In the next few sections, we discuss the detail of SCANRAW. It includes general processing, the architecture and each operator.

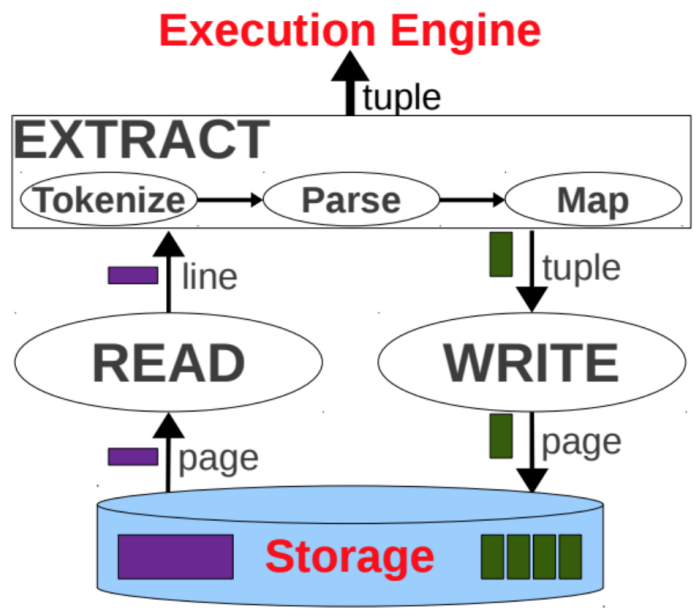

Figure 3.1 The general processing for SCANRAW

Figure 3.1 is query processing for the raw file.

The READ part represents scanning rows in a page from the flat file. Rows are read one by one and passed to the EXTRACT stage. We use a feature functionality to improve our performance; that is, rows that are stored in the same page on disk can be loaded into a batch efficiently; this is implemented by the File System. In addition, the other functionality we use is to cache rows, and the next time, we can retrieve them in the memory rather than the disk. This is also implemented by the File System.

The EXTRACT part contains three stages: TOKENIZE, PARSE and MAP. The TOKENIZE part is in charge of identifying each column of the tuple. The outcome of TOKENIZE is a set of starting offsets for columns of a tuple, which is passed to the next step, i.e. PARSE. The implementation of TOKENIZE is very straightforward; the idea is to go over each character in one text line, extract one column once the pointer encounters the delimiter or reach the end of the line by identifying '\n', and then put it into the set of starting offsets. There seems to be no optimization at this point. In some cases, it can be optimized; for example, people only need some of columns, not entire columns. Therefore, TOKENIZE only needs to scan specific columns, such as the first and third columns. TOKENIZE can skip the rest of the columns. There is a significant improvement if we need only a few columns that are located at the front of a tuple, and TOKENIZE can thus stop scanning once it reaches the last column we need for a given query and jump to the beginning of the next text line. All of the starting offsets are stored in a set, which we called a positional map.

In the PARSE stage, an echo column we extract from the raw file is converted into the binary format according to the data type. For example, the string 123 is converted into the integer of value 123. Some scenarios that can speed up the processing: one is if the number of columns for the given query is smaller enough compared to the number of all columns. Moreover, we push down selection into the scan operator. This helps us to enhance the throughput of reading since some of the rows are filtered by the predicates. Another technique we can use is the Cache. We cache all the converted column values in the main memory such that they can be reused. These techniques allow us to reduce the parsing time.

The last stage is MAP. The motivation of MAP is to organize the converted column values to the format that can written into the disk. For example, if we use the row-based format to store the data in the disk, then we organize them as a row and put them into the MAP. By contrast, if we employ the column-based format, then we organize the data as a column and put them into the MAP. Once the tuple we need is in the MAP, we do not use TOKENIZE and PARSE modules.

The WRITE module puts the binary data in the MAP into the disk. This processing is highly likely loading in the database-manager system. The cost of processing a query is due to the space in the disk and the I/O time. There could be a concurrent issue if one cannot handle the READ and WRITE correctly. This issue is not in sequential processing. The raw data are read, tokenized and parsed in the main memory, converted into the binary format, and put into the disk. This is called the READ-EXTRACT-WRITE pattern, which guarantees that there is no overlapping during the procedure. Another processing pattern that most of external tables use is READ-EXTRACT-LOAD-PROCESS

In the SCANRAW, we implement a dynamical mechanism that can determine a processing pattern that is used in the runtime. When the I/O is ideal, SCANRAW changes the pattern to partial data loading, take advantage of overlapping and processes them and loads/writes them into a disk/DBMS.

## 3.2 SCANRAW Architecture

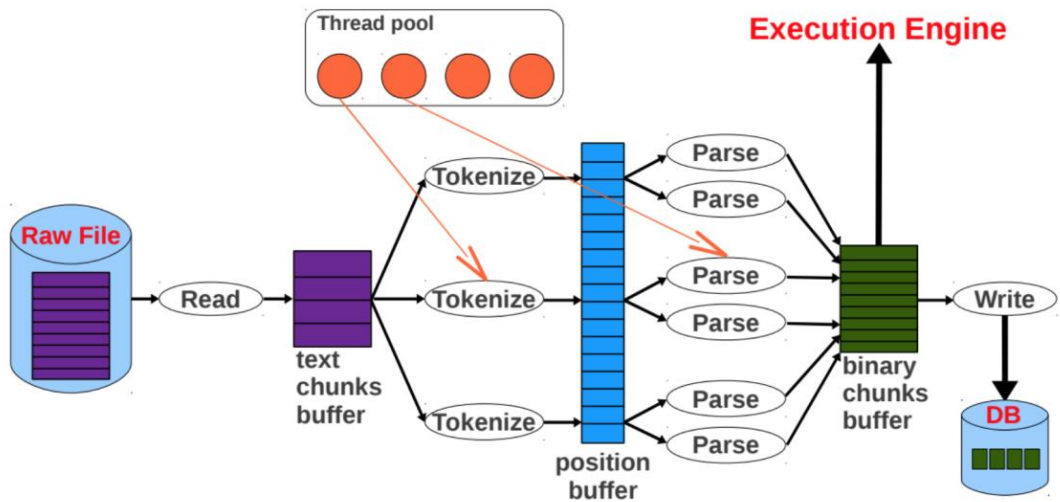

Figure 3.2 The architecture of SCANRAW

Figure 3.2 illustrates the architecture of SCANRAW. As can be seen, the figure is similar to the query-processing figure. However, it has more detail about how to parallel them.

In Figure 3.2, TOKENIZE and PARSE are scheduled by the thread pool. The scheduler can assign works to them during the runtime. READ and WRITE are also controlled by the scheduler. The difference in the scheduler's strategy can be the processing pattern. For example, it can be an external-table pattern if the scheduler does not call WRITE to persist binary data into disk/DBMS. It can also be Extract-Transform-Load (ETL) if the scheduler invokes WRITE for every tuple the READ get. For our system, we use a different policy, which is speculative loading. The idea is to trigger WRITE when the READ buffer is full. The strategy allows us to minimize the execution time but not maximize the I/O throughput.

In the figure, the structure of the super-scalar pipeline can be determined at runtime or be static at compiling time. The reason why it is implemented in this manner is that the optimal pipeline structure can be different in terms of a given data set. For example, if all or most columns have a numeric data type, such as Integer and Double, then the most expressive part would be the PARSE stage. If all or most columns have a string data type, then converting a string from the raw file to string type is very affordable. Only at runtime can we know the data type, and the dynamical pattern can give us much more optimizations. When the text-chunk buffer and the position buffer have some tuples, the scheduler would assign workers from the thread pool to each operator. The maximum of each operator depends on the capacity of the thread pool. The worker cannot execute the job until the thread pool has an available thread. In addition, the number of threads is configurable and can be dynamically changed.

We manage the buffer modules, which include the text-chunk buffer, the position buffer and the binary-chunk buffer, by using the producer–consumer design pattern. The size of each buffer is determined by the size of the main memory. The text-chunk buffer contains text tuples from the raw file. The file is stored in chunks, which is a bunch of lines. The processing happens in chucks at a time. The position buffer includes the text chunk and its corresponding positional map. The binary-chunk buffer contains the binary format of chunks, which are the final representations in the disk/DBMS. In this binary format, tuples with some or all of the columns are stored in the column base. Each column is written into the disk by a set of pages.

The binary-chunk buffer has the functionality of cache if the similar data used for the given query. The different policy of binary chunks buffer can make SCANRAW different. If the policy of the binary-chunk buffer is to cache all of the tuples, then the DBMS behind SCANRAW can be an in-memory database. In addition, the policy of binary-chunk buffer could be to use Least Recently Used (LRU) to replace tuples.

Moreover, the SCANRAW can pre-fetch chunks into the binary-chunk buffer. When the buffer is full, and chunks are not ready to be processed, pre-fetching stops. Chunks that are brought by pre-fetching can be replaced by the policy of cache. They can all be stored in the memory or dropped partially. Pre-fetching can be applied in the text-chunk buffer or the binary-chunk buffer. Furthermore, the binary-chunk buffer plays a crucial role of synchronizing SCANRAW with the execution engine.

## 3.3 Operation

In the section, we introduce the whole architecture of SCANRAW. We discuss how each operator (READ, WRITE, TOKENIZE, PARSE and SCHEDULER) coordinates with others. As Figure 3.3 illustrates, operators extract worker threads from the thread pool and configure threads with different tasks. For example, the READ operator obtains a worker thread and assigns a specific task to the worker thread. Next, we present each operator.

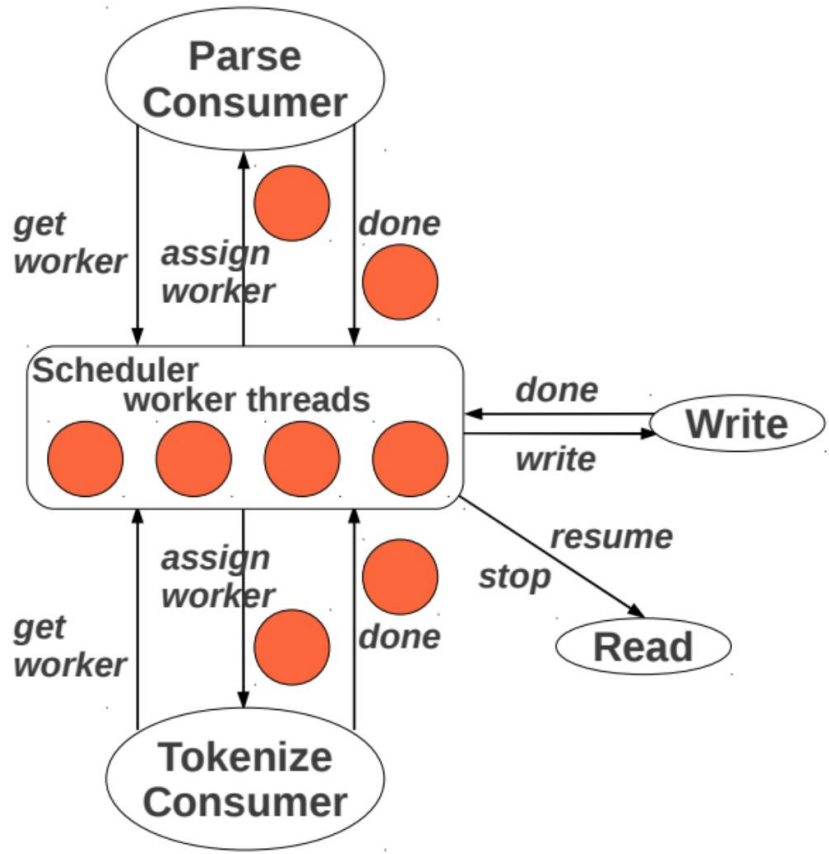

Figure 3.3 Operators and their control mechanism

The READ thread scans a bunch of rows (chunks) from the raw file and stores them in the text-chunk buffer. READ can look through the text chunks when the text-chunk buffer has available rooms. It will stop scanning when the buffer is full and thus needs to be consumed. Besides, the behavior of READ can be affected by SCHEDULER, which can allow or stop READ from checking the raw chunk data to avoid disk interference with the WRITE operator.

There are some optimizations for READ. Chunks can be read in a non-sequential order. In addition, the operator can be ignored if all rows in chunks do not satisfy the predicates. Moreover, chunks in the text-chunk buffer (already in the memory) can be used instantly. Furthermore, binary data in the binary-chunk buffer can also be used without tokenizing and parsing again. In the whole picture, we have three tiers for READ: the first is cache chunks, the second layer is chunks inside the DBMS and the final one is scanning from the raw file.

The WRITE thread is in charge of stocking the binary chunks into database. WRITE retrieves chunks from the binary-chunk buffer and materializes them into the DBMS. To achieve the highest throughput, SCANRAW only allows either READ or WRITE to use the disk at a time. Therefore, there is no interference for I/O and can gain the good performance.

The consumer thread audits every buffer, such as the text-chunk buffer, the positional buffer, and the binary-chunk buffer. The consumer thread typically picks up a worker thread from the thread pool, and the worker thread transfers the available buffer from one tiers of a buffer to the next. For example, the PARSE consumer thread sends a request when the next chunk is ready to be parsed. When a worker thread is assigned, the chunk we would like to process is selected and sent for processing.

The scheduler thread is responsible for managing the thread pool and answering a request from the consumer thread. When the thread pool has an available worker thread, the scheduler chooses one of threads and gives it to the consumer thread. One thing to which we should pay attention is that SCANRAW needs to check if there is available chunk buffer for the consumer thread. The reason for this is that if there is no available chunk buffer, a consumer thread cannot be executed because there is no space to allow the worker thread to put chunks from one place to another, even if the consumer thread acquires the worker thread.

The worker threads are always static. Each task for the workers is implemented before runtime. During the runtime, the worker thread is assigned one or some of tasks into it, and the worker thread executes them using their own CPU and memory resources.

# Chapter 4

# Code Generation for Raw Data Processing

## 4.1 Introduction

In this chapter, we propose a novel idea that combines code generation and query processing over an external table. The problem we would like to solve is how to apply code generation into query processing with raw data. Since most of modern machines have enough main memory, people can now buy 1 TB of DRAM for approximately $50,000, which is still decreasing. Moreover, one can have a fast disk or even a solid state drive (SSD) or a Non-Volatile Memory Host Controller Interface Specification (NVMe). This storage equipment performs excellent work. For example, Samsung 950 NVMe PCIe can achieve a throughput of 1,570 MB/s for reading and 763 MB/s for writing, which are significantly fast.

The prices for this fast storage are not too expensive any more. Many personal computers or laptops currently use SSD or NVMe as storage, which is common, and many companies have already moved the flash storage disk into their server machines. So far, we definitely can say the fast flash storage is popular worldwide, which means that the I/O is no longer the bottleneck of performance in terms of the size of space and the throughput.

Regarding code generation, the problem of I/O bound is solved by the hardware. Now, the computing resource is the most time-consuming part and CPU efficiency is crucial for performance.

The traditional program is that one composes some code, compiles it and executes it. The code one wrote should handle most of scenarios in a general manner. It is very common that there are many if-else conditions or switch statements inside the program. It significantly reduces the program's performance. The reason for this is that the if-else branch increases the instructions after compilation, and executing a program thus takes a longer time. This branch also removes the benefit of data locality. Since the program cannot determine the branch it should use during compiling time, it only can determine this at runtime. To interpret and execute the branch statement, the CPU removes old data inside the cache or registers and pulls new information that the branch needs into its cache. This process delays programs delay and is an expensive operation.

For example, the code for a simple for loop is illustrated in snippet 4.1.

**Snippet 4.1**

```
For i from 1 to 100

    A[i] += A[i - 1];

End
```

As you can see, the for loop calculates each item using its preceding item. The compiler is smart. When the code is executed, the CPU knows that the next iterator must have the previous item and the current item, meaning that each iterator is predictable. So, the CPU can put some data that must be used in the next iterator inside a cache or register; it is definitely fast.

We compared snippet 4.1 with the code below (snippet 4.2).

**Snippet 4.2**

```
For i from 1 to 100

    A[i] += A[100 - i - 1];

End
```

The only difference in snippet 4.2 is that we change the logic slightly. In each iterator, the current item adds the last $100 - i - 1$ item. In this scenario, which is more complicated, the CPU cannot predict the next iterator; it has to figure it out once it is in the iterator. In snippet 4.2, the CPU had to kick out the data that should be used in the next iterator and reload the data it needed. There are many times data are transmitted back and forth since the CPU totally does not know data that should be used in the next step.

Our solution, i.e. generating a code with an invariant and unrolling the for loop as possible as we can, has fixed this problem.

Another technique we use to improve the performance is reducing the number of branches. For example, people write a program that typically covers all the scenarios as much as possible; the practical approach to handling a code that processes different scenarios is to use if-else conditions or switch statements. For example, when a query comes in, the query engine generates a plan and executes the plan over the data inside the database. The most work the execution engine does is to check each data type of a tuple and cast every column of each tuple to the original data type. One may think the following snippet of the code is very common in the database-manager system.

**Snippet 4.3**

```
For each col in each tuple

    object = read(tuple, offset);

    if(instance of object == INT) {

        cast it to INT;

    }

    else if(instance of object == DOUBLE) {

        cast it to DOUBLE;

    }

    else if(instance of object == STRING) {

        cast it into STRING;

    }

    ………

End
```

As can be seen, the above code (snippet 4.3) has a number of if-else conditions; if we would like to cast more data types, then we would write more if-else conditions, whose number equals the number of data types.

Technically, the if-else condition and switch statements are not a CPU friendly. They can increase the instructions of a machine code and break the pipeline of a CPU such that the CPU cannot determine the instruction that should be executed in the next stage. In addition, they bring many data locality issues. The pennies are very expensive once the data is kicked out from the cache/register of CPU.

To build a database for general user cases, for example, many companies use the tuple-at-a-time or volcano model. However, those kinds of model use many virtual function calls to invoke the next operator. The virtual function, as we discussed in the previous chapter, needs a vtable and can be interpreted and cast to an original function.

Those overheads can be eliminated by using code generation. For branches, we can get enough information to choose one of them as the target branch and discard the rest, since they are not used in the current scenario. Additionally, with using code generation we can easily generate functions and snippets of a code without scanning the virtual function.

## 4.2   Design

In this section, we present our design for processing a query through code generation. There are several models involved. The first one is the SQL parser, through which we define some SQL pattern statements that include select, load, join, group by, distinct, etc. These are very common SQL query statements. Furthermore, we design some data structures according to these SQL query statements. For example, the select statement needs a select list and predicates; we design the select list and predicates to be linked lists. When a query comes in, it is parsed and tokenized into different data structures. Those data structures can be employed in the later stage. At the same time, the checking function performs. For instance, a query is scanned, and if it has some table with invalid columns (i.e. it does not exist in the catalog or metadata), then the query is labeled as an invalid query. None of models in the next stages will take it.

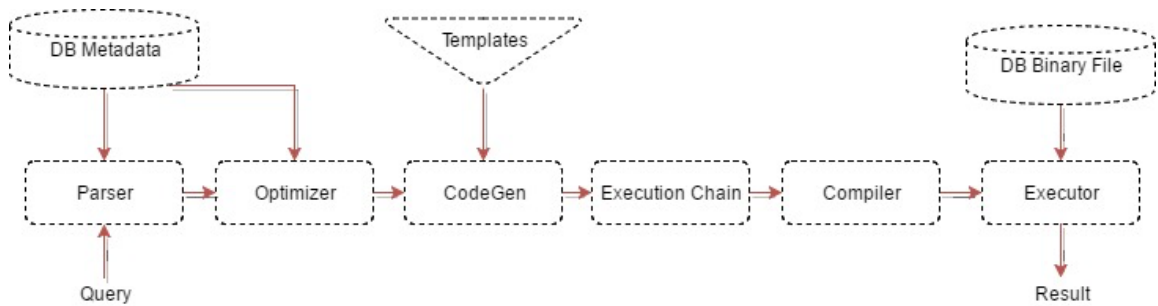

Figure 4.1. The overview of a query engine

In the second phase, if the query is passed and valid, join tables inside the query are processed. First, the Optimizer pull some information, such as the number of distinct items and the number of tuples in each table. That information helps us to determine which join order is the best optimization.

There are two main optimization algorithms or solutions. One is the left deep tree or the right deep tree in some other database-manager systems. This solution applies the greedy algorithm to find the join order. Each time, it only chooses the cheapest join pair and then uses the left one because the left one is always a fixed side or hash side (using a hash join). By contrast, the right side is always the probe side. In addition, the other algorithm for computing an optimization tree is the permutation algorithm, which is based on the Dynamic Programming. Essentially, it goes over every permutation of the join order and finds the most affordable one as the final result. In addition, one can slightly enhance the dynamic programming algorithm through the technique of cutting branches. The algorithm can reduce some of unnecessary cases, thereby saving some time during the runtime.

When the optimized tree or join order is done, the third section is code generation. We choose to use the template approach to implement code generation, which uses GNU M4 as the tool. In addition, we add code fusion into our work, which is mentioned in HyPer. Thus, we can achieve higher performance. The benefit from HyPer or code fusion is that it improves data locality. We discuss it in detail in the next section.

In the fourth stage, we build the execution chain; in fact, it is the topologic chain. The execution chain includes the rest of operations beside the join operation(s) since the join tables are already covered in the optimization part.

In the fifth phase, the generated code is compiled with GCC or Clang during the runtime. The generated object should be the shared object. Then. the code is loaded into program by using system.loadlib.

In the final part, the executor can execute the query with the specific code we generated, which is faster than the general one.

## 4.3  Template Approach

The template approach for code generation is one of most popular text-based code-generation processes. The idea of this approach is to use some metadata as a controller that is given by the query that comes in and fill out a text-based template. There are some placeholders that are filled by the information from the controller during the runtime.

There are three roles for a text-based template of code generation: the input, the transformation and the output. It is similar to MVC. Model–view–controller (MVC) is a software architectural pattern for implementing user interfaces on computers. It divides a given application into three interconnected parts and can render the final view based on the given input.

There are many ways to generate a code using the template approach. One of them is direct string concatenation (also known as the Brute Force Approach). It is the earliest approach that has always been straightforward. For example, if we would like to generate a select operator for the name John, the direct string concatenation scans and collects all the items whose name is John.

**Snippet 4.4**

```
string cppcode;

cppcode += " #include<iostream>";

cppcode += " #include<vector>";

cppcode += " using namespace std;";

cppcode += " Tuple select(table r1) {";

cppcode += " foreach(tuple one : r1) {";

cppcode += " if (one.name = " + $condition " ) {";

cppcode += " HashTable.put(one)";

cppcode += " }"

cppcode += " }";

cppcode += " }";

FileWriter.write(cppcode);

FileWrite.flush();

FileWrite.syn();

FileWrite.close();
```

As can be seen in Snippet 4.4, the placeholder named $codition is replaced by the string "John", and the C++ code that is customized to scan only John's tuples can be generated.

This implementation of the template approach is easy to use. It is hard, however, to apply this idea to a more complicated scenario. The functionality is not only limited but but also faces other issues, such as Character escaping, Character encoding as well as managing and reusing templates.

Therefore, many template engines come out to handle complex template-based demands, such as CodeDom, Microsoft T4, Apache Velocity and GNU M4. These are very powerful tools that are designed to allow users to write an efficient and complicated generated code. In addition, they support many built-in features of which one can take advantage, for instance, generating multiple files from one template.

In our scenario, the intention of using code generation with the template approach is to customize the C++ program and convert the general one to a specific one for a particular query, and the final ad-hoc program should also be written in C++. For the transmission part, in our case, we use GUN M4 as the transmission language. GUN M4 is a tool, as it is an implementation of the traditional Unix macro-processor. It supports many useful template features, such as a variable, the for each loop, the if-else condition, multi-byte characters, function definition, multiple arguments and joining multiple text arguments. These features are good enough for our code-generation scenario.

In addition, we would like to use the HyPer approach and combine it with the template approach.

## 4.4  HyPer Approach

To maximize the performance of query processing, besides the template approach we have used, the idea of data locality would be applied into our work.

There are terms we would like to introduce that are defines in the HyPer approach. The pipeline breaker is an algebraic operator for a given input side, if it takes an incoming tuple out of CPU registers. A full pipeline breaker materializes all incoming tuples from the build side before continuing with processing.

The motivation for us to use the HyPer approach is to keep all data in the CPU cache as long as possible during query processing. Thus, we can take advantage of data locality to improve the performance. The tuple-at-a-time model is inappropriate for the data locality. One of the reasons for this is that each function call in the volcano model is a virtual function call, which always the removal of data out of the cache inside of the CPU. Some databases, such as Impala, use batch tuples at a time, and they can amortize time complexity. However, it still cannot avoid removal of data from the CPU registers. Actually, any kind of architecture that is based on the volcano model, which pulls tuple(s) up from one operator to another, can break the pipeline and is not a data-locality paradigm.

Another reason why it is hard for the iterator model to enhance performance is that it blends the control flow and data flow. As mentioned earlier, in the tuple-at-a-time model, each tuple is pulled from one operator to another, and the path it goes through is the path of control flow, which means that a tuple must finish a control logic flow, which puts much pressure on the CPU caches.

## 4.5  Implementation

In the previous section, we described the template approach and the HyPer approach. Now, let us introduce how we implement it.

To compare a non-HyPer approach with the HyPer approach, we implement two versions of code based on the template approach. In other words, we use a positional map to connect all operators in the code-generation phase for the non-HyPer approach, and we use code fusion to combine operators together inside the same for loop to achieve data locality, which is mentioned in the previous sections.

The template approach without code fusion

In our scenario, we have to first parse and tokenize the raw data, which are stored in a CSV file. CSV is a simple file format; each column of a CSV file uses a comma to separate it from others inside one tuple.

How does each column read each part of data in a tuple or row? Suppose the separator is a comma (,), we read each line from its beginning to its end. When we encounter a comma, which means we have already read column data, we begin to read the next adjacent column. We employ the template approach at this point, which allows us to easily get each data for each column since the order of the columns is fixed; thereafter, the straightforward way is to use a for loop with an index of the column in the schema of tuples.

**Snippet 4.5**

    For each column
        Init Data
                While(! read encounter comma)
                        Data.Append(byte)
            Assign Data to column

As can be seen in Snippet 4.5 above, the for-loop statement, as we mentioned in the previous chapters, can lead to jump instructions and that break the pipeline and prediction of the CPU. The program can be slower since it uses jump instructions, and it is difficult for the compiler to optimize the code during the compilation time.

By using the template approach, we can unroll the for loop and eliminate jump instructions. In addition, the compiler already knows the behavior of the code, so it is advisable to use some features to optimize the code by using a compiler.

**Snippet 4.6**

    M4_foreach(</_C_/>, </COLUMN_TYPES/>, </dnl
    Each column
        Init Data
                While(! read encounter comma)
                        Data.Append(byte)
            Assign Data to column
    End of M4

Snippet 4.6 illustrates how to unroll the for loop using M4. M4_foreach is a macro language that can present all the columns separately with the same pattern after the M4 engine compiler, as illustrated by Snippet 4.7.

**Snippet 4.7**

```
Col1
    Init DataCol1

            While(! read encounter comma)
                    DataCol1.Append(byte)
        Assign Data to Col1
Col2
    Init DataCol2

            While(! read encounter comma)
                    DataCol2.Append(byte)
        Assign Data to Col2
Col3
    Init DataCol3

            While(! read encounter comma)
                    DataCol3.Append(byte)
        Assign Data to Col3

……
```

We can view the tokenizing part as an operator; then, the next operator can fill the data into an array that is associated with the column. To fill the data, we can still use the template approach. In our case, to fill the data, we need to put the data into the variable of an array whose name corresponds to the column. For instance, the column ID can be put into the array named id_array; each index inside id_array represents the order of tuples. In C++, it is impossible to concatenate a column name and a type name to form an array name. However, using the M4 template approach, one can easily replace and concatenate any name of an array or a variable.

However, how can one connect the tokenizing operator and the filling operator or any other two operators? We employ a positional map as the medium to connect two parts. The positional map records a column and its corresponding pointer in each tuple. One can apply this information in the next operation.

**Snippet 4.8**

```
m4_foreach(</_C_/>, </COLUMN_TYPES/>, </dnl
    // tokenizing
    curr=next;
    while ( (next-buffer) < MAX_LINE_SIZE && *next != 'SEPARATOR' &&
        *next!='\n') next++;
    *next = 0;
    next++;
    positional_map[i].push_back(make_pair("M4_COL_NAME(_C_)", curr));
/>)
dnl # END OF M4 CODE

……

for (int i = 0; i < _tuplesNo; i++) {
  if(positional_map.find(i) != positional_map.end()) {
        for(auto it = positional_map[i].begin(); it != positional_map[i].end();
  ++it) {
    m4_foreach(</_C_/>, </COLUMN_TYPES/>, </dnl
      if(needs.find("M4_COL_NAME(_C_)")!=needs.end()) {
        dnl #if(it->first == "M4_COL_NAME(_C_)") {
        M4_COL_FROM_TOKEN_ARRAY(_C_, i, it->second)dnl
        M4_COL_MINMAX_ARRAY(_C_, i)dnl
        }
    />)
    }
  }
}
```

As the snippet code 4.8 illustrates, the positional map can share the pointer within operators.

In the template approach with Code fusion, there are similar steps between the template approach without code fusion and the one with code fusion. The main difference is that two operators can be merged into one for loop. The essential idea of HyPer is to fuse the code to enhance data locality. When we process the tokenizing part, each column of the data can be written into the slot of an array in the same iterator; this is a very good practice of pipelining. We can keep it in the CPU register and materialize it into the memory.

**Snippet 4.9**

```
  m4_foreach(</_C_/>, </COLUMN_TYPES/>, </dnl
     // tokenizing
     curr=next;
     while ( (next-buffer) < MAX_LINE_SIZE && *next != 'SEPARATOR' &&
         *next!='\n') next++;
     *next = 0;
     next++;
   />)
   dnl # END OF M4 CODE
 ……
   for (int i = 0; i < _tuplesNo; i++) {
           for(auto it = positional_map[i].begin(); it != positional_map[i].end();
     ++it) {
       m4_foreach(</_C_/>, </COLUMN_TYPES/>, </dnl
         if(needs.find("M4_COL_NAME(_C_)")!=needs.end()) {
           dnl #if(it->first == "M4_COL_NAME(_C_)") {
           M4_COL_FROM_TOKEN_ARRAY(_C_, i, it->second)dnl
           M4_COL_MINMAX_ARRAY(_C_, i)dnl
           }
       />)
     }
   }
```

The code Snippet 4.9 demonstrates that we put two operations together, which are the tokenizing operator and the filling column operator. In each iteration, when we get the current data pointer and know how long the column is, we can immediately write this piece of data into the correspondent slot in the array, which can take advantage of data locality since the data are still in the register inside the CPU, and we would not kick them out. We keep them in the CPU and use them for the next operation. This helps us to get a higher performance and throughput.

## 4.6 Experiments

To test the performance of code generation, we use DBGen to generate different sizes of TPCH data with different factors. To evaluate query processing, we choose the LINEITEM table as the test relation.

We measure the cost time on the lab master cluster, which uses 16 CPUs, each of which has 8 cores. The model of CPU is AMD Opteron(tm) Processor 6128, which clocks at 2.0 GHz. The cache size is 512 KB, the size of DRAM is 65 GB, and the operating system is Ubuntu 14.04.5 LTS.

To measure the cost time, we use std::clock_t to mark the start time and end time and then calculate the running time by subtracting the start time from the end time. However, there are some factors that can affect the cost time if we use a time command to determine the cost time. For example, the compilation time of code generation can be seconds for the first compilation.

We categorize different scenarios for testing the performance and compare their throughout. The idea is that we design the query (load) data from a raw CSV file, i.e. LINEITEM, with distinct number of columns we would like to extract. For example, we can only load one column from the LINEITEM CSV file, or one can load half or all of them. Thus, we can compare the different throughputs with and without code fusion.

In addition, we also would like to test both functionalities in different data sizes as our second dimension of experiments. For instance, one can do an experiment on 1 GB data and 7 GB data to see how data locality enhances performance.

The Figure 4.1 indicates that the performance of the query processing for one column with code fusion is much faster than that of the same method without code fusion. As one can see, the gap between both is large, i.e. at most 3 seconds.

For this test case, we choose the column l_orderkey as the column we would like to extract from the CSV raw file. The data type of l_orderkey is int, meaning that it takes 4 bytes to represent itself. The whole tuple would be around 200 bytes, including the string data type that we estimate to be 100 bytes. So the orderkey column considers 4/200, which is 2% data. For example, if the raw file data size is 2.2 GB, then we can expect the size of loading data is 0.044 GB from the CSV raw file. The is highly optimized if we only care about the column of l_orderkey. The performance of code fusion with the template approach is faster since it keeps each piece of data in the CPU's cache and reuses the data from the register for the next operator. In addition, it does not need the positional map; thus, it saves the cost of storage of the hash table. The map in C++ is implemented by a red black tree, which means we also save the cost of operations through the red black tree.

Table 4.1 Loading one column

| Size of Data | WithCodeFusion | WithOutCodeFusion | Different | Diff in Percentage |
|---|---|---|---|---|
| 0.7 GB | 147.709 | 218.849 | 71.14 | 33% |
| 2.2 GB | 430.077 | 656.135 | 226.058 | 34% |
| 4.4 GB | 873.856 | 1331.62 | 457.764 | 34% |
| 7.3 GB | 1461.13 | 2219.71 | 758.58 | 34% |
| 11 GB | 2201.99 | 3343.55 | 1141.56 | 34% |
| 26 GB | 5324.41 | 8184.76 | 2860.35 | 35% |

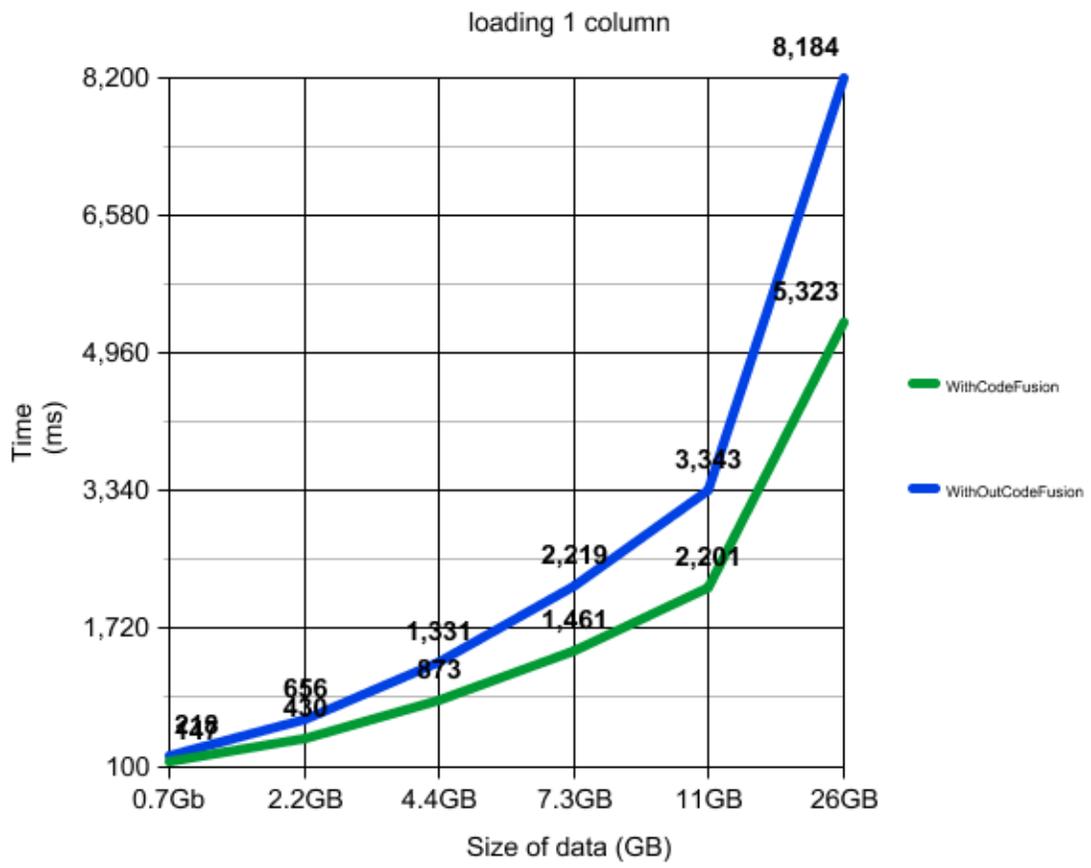

Figure 4.2 Loading one columns

In the next experiment, we load more than one column, i.e. 4 columns. As Table 2 and Figure 4.2 demonstrate, the method with code fusion is slightly faster than the one without code fusion. Compared to the experiment of loading one column, the difference between the two experiments is smaller than that in the previous test. It increased from 14% to 21%.

Table 4.2 Loading four columns

| Size of Data | WithCodeFusion | WithOutCodeFusion | Different | Diff in Percentage |
|---|---|---|---|---|
| 0.7 GB | 199.113 | 252.053 | 52.94 | 21% |
| 2.2 GB | 605.331 | 752.592 | 147.261 | 20% |
| 4.4 GB | 1248.05 | 1537.31 | 289.26 | 19% |
| 7.3 GB | 2209.28 | 2637.24 | 427.96 | 16% |
| 11 GB | 3262.11 | 3996.88 | 734.77 | 18% |
| 26 GB | 8420.86 | 9799.87 | 2860.35 | 14% |

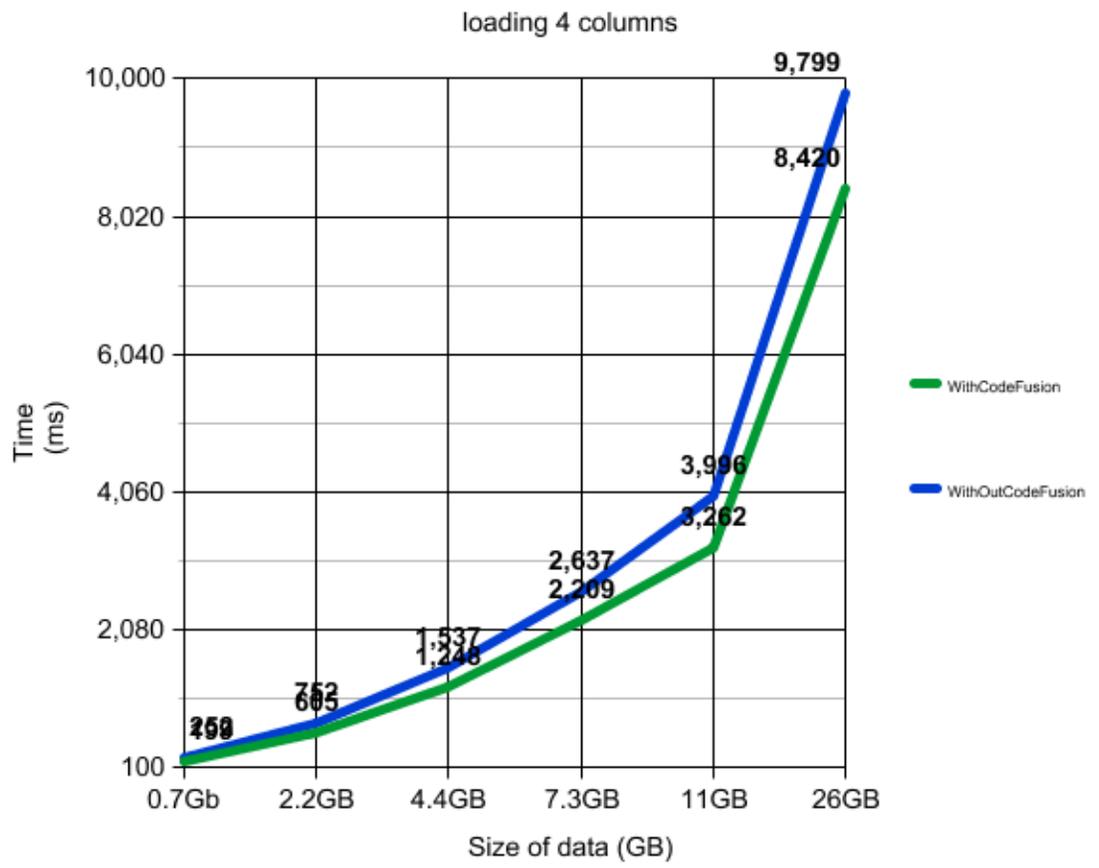

Figure 4.3 Loading four columns

In this experiment, we load eight columns which is half of all columns in table LINEITEM. In Table 3, one can see that the code-fusion version runs slightly faster than that without code fusion. Fig 4.3 also indicates the difference between the two versions is smaller than that in pervious experiments.

Table 4.3 Loading eight columns

| Size of Data | WithCodeFusion | WithOutCodeFusion | Different | Diff in Percentage |
| --- | --- | --- | --- | --- |
| 0.7 GB | 245.892 | 288.016 | 42.124 | 17% |
| 2.2 GB | 746.597 | 897.184 | 150.587 | 17% |
| 4.4 GB | 1545.98 | 1804.72 | 258.74 | 14% |
| 7.3 GB | 2611.18 | 3082 | 470.82 | 15% |
| 11 GB | 4083.21 | 4668.4 | 585.19 | 13% |

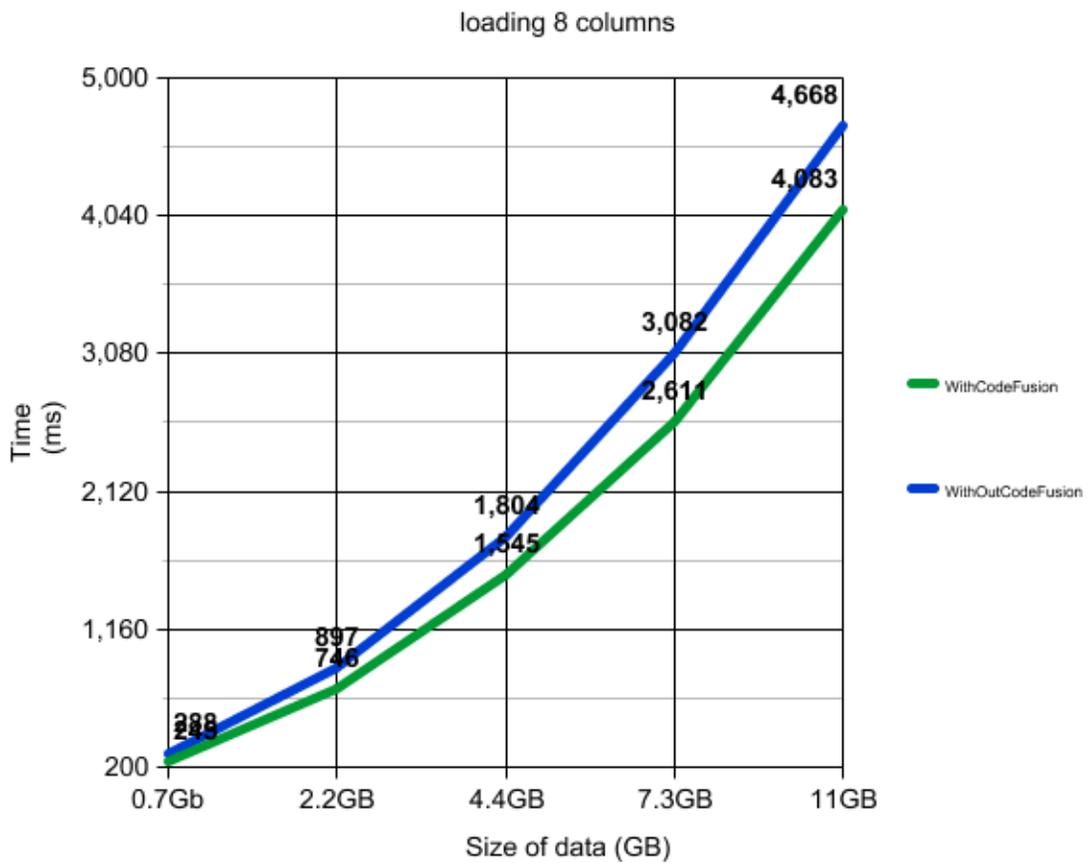

Figure 4.4 Loading eight columns

In the last experiment, we try to include all columns in table LINEITEM. In addition, the version with code fusion performs faster than the one without code fusion. However, the trend of the gap is decreasing; it is from 7% to 11%. This range of the difference is the smallest one compared to all of the experiments we conducted previously.

Table 4.4 Loading entire columns

| Size of Data | WithCodeFusion | WithOutCodeFusion | Different | Diff in Percentage |
|---|---|---|---|---|
| 0.7 GB | 330.987 | 360.404 | 29.417 | 8% |
| 2.2 GB | 993.899 | 1121.83 | 127.931 | 11% |
| 4.4 GB | 2047.58 | 2249.73 | 202.15 | 9% |
| 7.3 GB | 3646.74 | 3870.67 | 470.82 | 6% |
| 11 GB | 5827.06 | 6239.8 | 412.74 | 7% |

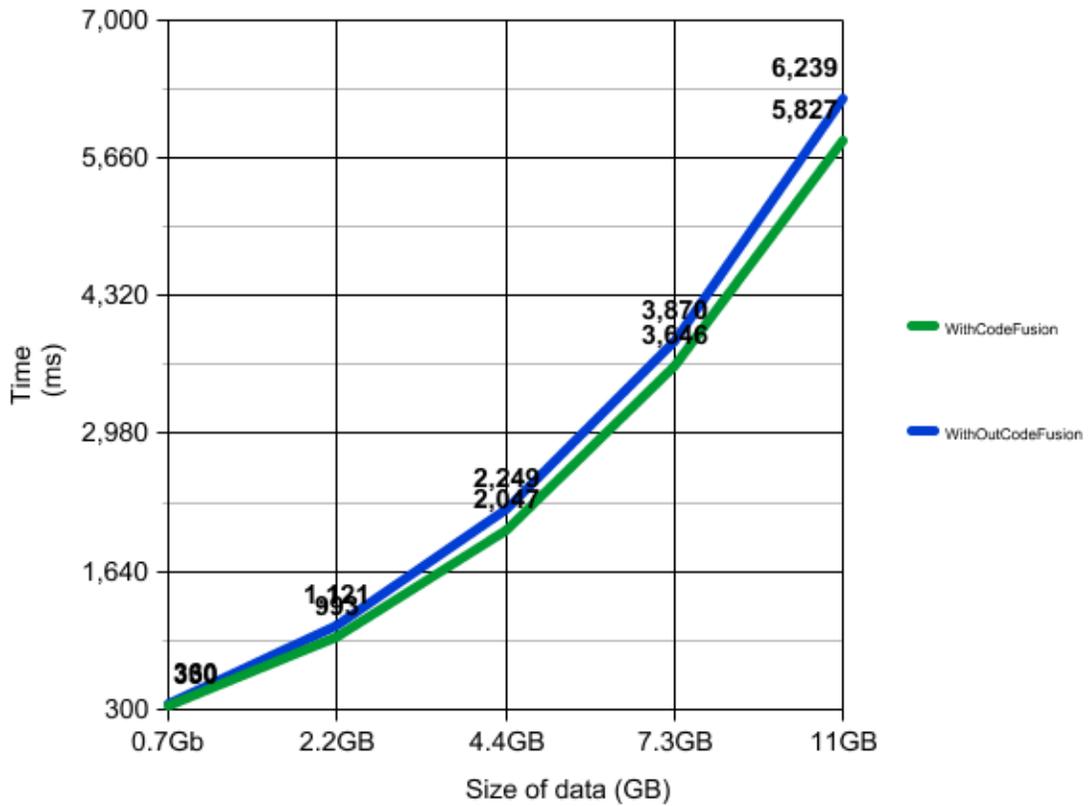

Figure 4.5 Loading entire columns

# CONCLUSION

In this paper, we have discussed the basic concepts of the DBMS and a query over a raw file. We propose the SCANRAW as well as its architecture, buffer mechanism and operations. There are many more innovations from SCANRAW. Reading data from the raw file on a disk is scanning in chunks at a time, which can take advantage of a File System. The system cache functionality can also speed up the process for us. In addition, multi-tiers of buffers are another method of optimization to reduce access time to the disk. The well-designed mechanism of SCANRAW threads and control messages brings an efficient I/O to its performance. Speculative loading is one of major novel ideas in SCANRAW. The motivation of SCANRAW is to find time intervals during query processing when there is no reading from the disk. It is proved that query processing is CPU-bound. However, the I/O is an idea the time when speculative loading happens.

In the code-generation section, we also introduce different approaches for code generation, such as the template approach, bees and HyPer code fusion. The main idea of the template approach is to find the part of code that is determined at runtime and leave this part as a blank space that is filled out during runtime. The bees approach is a more likely optimized invariant and focuses on the function level, not the whole class level. The HyPer approach is maximizing data locality and efficiently uses the iterator in the for loop between multiple operators. Another HyPer approach is based on the original HyPer approach that incorporates SIMD and Prefetching techniques. Additionally, we design and implement our own code for query processing based on the template approach with code fusion.

The experiments we conducted were performed on the master node of the GLADE cluster. The benchmark data are generated by DBgen, which contains TPC-H data. Specifically, we used LINEITEM as the sample input raw-file data. Experiments have proved through comparison the difference in performance between the template approach with code fusion and the one without code fusion. Our approach indicates that if we only need one or some parts of columns. not all columns, the performance with code fusion can be faster and more efficient and avoid unnecessary reading and loading.